\documentstyle[aps,prb,multicol,axodraw1]{revtex}
\input epsf
\renewcommand{\Im}{\mbox{Im}}
\renewcommand{\Re}{\mbox{Re}}
\renewcommand{\vec}[1]{{\bf{#1}}}
\newcommand{\bra}[1]{\langle {#1}  \! \mid}
\newcommand{\ket}[1]{\mid  \!  {#1} \rangle}
\newcommand{\gvec}[1]{\mbox{\boldmath $ #1 $\unboldmath }\! }
\newcommand{\ts}{\textstyle }
\newcommand{\lsim}{ {{}_{\ts <} \atop {\ts \sim}} }
\newcommand{\gsim}{ {{}_{\ts >} \atop {\ts \sim}} }
\newcommand{\lrule}{ \end{multicols} \noindent
  \rule{0.5\textwidth}{0.1mm}\rule{0.1mm}{3pt}\newline }
\newcommand{\rrule}{ \noindent \parbox{\textwidth}{
  \hfill\rule[-3pt]{0.1mm}{3pt}\rule{0.5\textwidth}{0.1mm}} 
  \begin{multicols}{2} }
\begin{document}
\draft
\title{Effects of Strong Magnetic Fields on Pairing 
Fluctuations in \\ High Temperature Superconductors}
\author{M. Eschrig$^\ast$, D. Rainer$^\dagger $, and J.A. Sauls$^\ast$ }
\address{
$^\ast $Department of Physics \& Astronomy, Northwestern University, Evanston, IL 60208, USA\\
$^\dagger $Physikalisches Institut, Universit\"at Bayreuth, D-95440 Bayreuth, Germany}
\date{Received 15 December 1998}
\maketitle
\begin{abstract}
We present the theory for the effects of 
superconducting pairing fluctuations on the
nuclear spin-lattice relaxation rate, $1/T_1$,
and the NMR Knight shift for layered superconductors
in high magnetic fields. These results can be used to 
clarify the origin of the pseudogap in high-$T_c$ 
cuprates, which has been attributed to
spin fluctuations as well as pairing fluctuations.
We present theoretical results for s-wave and
d-wave pairing fluctuations and show that recent 
experiments in optimally doped
YBa$_2$Cu$_3$O$_{7-\delta}$ are described 
by d-wave pairing fluctuations.\cite{mitrovic98,bachman98}
In addition, we show that the
orthorhombic distortion in YBa$_2$Cu$_3$O$_{7-\delta}$ accounts for
an experimentally observed discrepancy between $1/T_1$ obtained by NQR and NMR
at low field. We propose an NMR experiment
to distinguish a fluctuating s-wave order parameter 
from a fluctuating strongly anisotropic order parameter,
which may be applied to the system Nd$_{2-x}$Ce$_x$CuO$_{4-\delta}$
and possibly other layered superconductors.
\end{abstract}
\pacs{PACS numbers:  74.25.Nf, 74.40.+k, 74.25.Ha}
\vspace{-11pt}
\begin{multicols}{2}
\section{Introduction}
Fluctuations are enhanced in high-$T_c$ cuprate 
superconductors because of their layered structure and their small
coherence length.\cite{fisher91} 
In contrast to
conventional superconductors, where the transition is very
well described by a mean field theory, an extended region of one to
several Kelvin around the transition is expected to be
dominated by critical fluctuations in the cuprates. 
In this paper we discuss the effects of
gaussian dynamical fluctuations above $T_c$, which 
are observable over a temperature
range $T-T_c \approx T_c$, on the nuclear spin-lattice
relaxation rate and the NMR Knight shift in high-$T_c$
superconductors.
For a comprehensive review on the role of
NMR-NQR spectroscopy in the study of fluctuation effects in high-$T_c$
superconductors see Rigamonti {\it et al}.\cite{rigamonti98}

Pairing fluctuation effects on the spin-lattice relaxation
rate have been investigated
in the dirty limit for static, long-wavelength fluctuations 
near $T_c$ by Kuboki and Fukuyama.\cite{kuboki89} 
Heym extended these calculations for s-wave pairing fluctuations
by including the fluctuation corrections to the quasiparticle
density of states.\cite{heym92} 
Analytic expressions for the static, long-wavelength fluctuation
corrections to the spin-lattice relaxation rate and Knight shift were obtained
by Randeria and Varlamov for ultra-clean and dirty s-wave
superconductors.\cite{randeria94}
We extend their calculations to include
finite magnetic fields and
unconventional pairing for general values of the 
impurity scattering rate. Our calculations and numerical results
include dynamical fluctuations and short wavelength fluctuations
summed over all Landau levels.

Dynamical quantities such as the fluctuation contribution to the
spin lattice relaxation rate, $1/T_1$, 
carry valuable information on the type of 
fluctuations and characteristic scattering rates and lifetimes.
Qualitatively different behavior for the fluctuation contributions to
the rate is predicted
for different symmetries of the order parameter fluctuations.\cite{carretta96}
Analysis of the fluctuation corrections to
$1/T_1$ provides information on the 
elastic and inelastic scattering parameters.
The {\it sign} of the fluctuation corrections to $1/T_1$
is sensitive to pair breaking and the
symmetry of the pairing fluctuations; thus, non-magnetic impurities
have almost no pairbreaking effect on fluctuations
with s-wave symmetry, but have strong effects for d-wave pairing.
In the case of s-wave pairing a large positive fluctuation contribution
to $1/T_1$
originates from the anomalous Maki-Thompson (MT)
process.\cite{CarMakThomp67,kuboki89}
We show that this 
process is suppressed in zero field almost completely
for d-wave pairing if the mean free
path is shorter than 20 coherence lengths, but cannot be neglected
near the transition in finite magnetic fields or in the ultra-clean limit. 
Fluctuation corrections to the quasiparticle density of states (DOS)
dominate the anomalous Maki-Thompson processes in the case of 
d-wave pairing symmetry
for realistic scattering parameters in high-$T_c$ cuprates.
For a recent review on the role of pairing fluctuation corrections
to the quasiparticle density of states in high-T$_c$ superconductors
see Varlamov {\it et al.}\cite{varlamov98}

Recent $^{63}$Cu NQR-NMR 
experiments on optimally doped YBCO 
by Carretta {\it et al.}\cite{carretta96} were interpreted in terms
of a pseudogap originating from superconducting fluctuations. 
Other theories for the pseudogap include spin-charge separation, 
preformed pairs, phase fluctuations, van Hove scenarios.
For a recent
review of this broad topic and references see Randeria.\cite{randeria97}
Chubukov {\it et al.}\cite{chubukov96}
proposed a magnetic mechanism for the pseudogap
in which ``hot'' quasiparticles become gapped by a precursor spin-density
wave.
Recent studies by Auler {\it et al.}\cite{aul97} 
of $^{63}$Cu and $^{89}$Y NMR in YBCO 
as a function of doping were interpreted as
evidence for the vanishing of the pseudogap for ``hot'' quasiparticles
due to antiferromagnetic spin fluctuations exactly at optimal doping,
whereas a pseudogap for ``cold'' quasiparticles
persisted at optimal and overdoped samples.
Whether the pseudogap is due to pairing fluctuations, spin-density wave
fluctuations, or more complicated mechanisms
may not be easy to decide, especially in optimally doped
materials. 
The study of fluctuation effects
in the presence of strong magnetic fields may be key to solving this
problem. 

Magnetic fields tend to enhance pairing fluctuations near
the transition temperature as a result of Landau quantization
of the orbital motion of pairs.\cite{aoi74} 
However, because the transition temperature
is suppressed by a magnetic field, pairing fluctuations are
typically reduced at constant temperature with increasing field.
Application of a magnetic field at constant temperature has very different
effects on the pairing fluctuation contributions to $1/T_1$
depending on the pairing symmetry.
For s-wave pairing the rate
is reduced with increasing field, whereas in d-wave pairing
the suppression of the DOS fluctuations, which have a negative
sign, leads to an enhancement of $1/T_1$ with field.\cite{carretta96}

In the next section we describe the theoretical 
framework for our analysis of fluctuation effects on NMR in high magnetic
fields in high-T$_c$ superconductors.
We derive the fluctuation propagator for a quasi-2D layered
superconductor and include quasiparticle scattering by non-magnetic impurities
in addition to pairbreaking by inelastic scattering. We incorporate orbital
quantization by the magnetic field on the pairing fluctuations as well as 
the effects of d-wave symmetry. In section \ref{nslr} we discuss the
pairing fluctuation corrections to the nuclear spin lattice relaxation
(NSLR) rate. To leading order in $T_c/E_f$ the dominant fluctuation corrections 
are determined by Maki-Thompson processes and corrections to the
quasiparticle density of states. We derive expressions for these processes
appropriate to 2D fluctuations in a strong magnetic field and
present our results for the pairing fluctuation corrections to the NSLR rate.
The field-dependence of the fluctuations is shown to be 
sensitive to the symmetry of the pairing fluctuations.
In section \ref{Susc} we derive the leading order corrections to the Pauli
spin susceptibility and its contribution to the Knight shift.
The Knight shift is determined by the the long-wavelength 
spin susceptibility, and in contrast to the NSLR rate the fluctuation 
corrections to the spin susceptibility are not very sensitive to order
parameter symmetry or impurity scattering. However, dynamical fluctuations
and orbital quantization lead to significant effects on both the rate and 
the spin susceptibility which are essential for a quantitative understanding 
of the pseudogap behavior in
high-T$_c$ cuprates. In sections \ref{nslr} and \ref{Susc} we
compare our theoretical results with
recent measurements of the pseudogap in the
NSLR rate and the Knight shift performed by 
Mitrovi{\'c} {\it et al.}\cite{mitrovic98} and Bachman {\it et al.}\cite{bachman98}
on optimally doped YBCO in magnetic fields up to 30 Tesla.
We show that the pseudogap in optimally doped
YBCO can be accounted for quantitatively by the theory of 2D pairing
fluctuations with d-wave symmetry.\cite{mitrovic98,bachman98}
Finally, we show that incorporating orthorhombic anisotropy
and the allowed mixing of s-wave and d-wave pairing fluctuation channels
leads to a low-field crossover from predominantly s-wave fluctuations
to predominantly d-wave fluctuations which provides a natural
explanation for the observed evolution from the NQR rate to the
low-field (below 2 Tesla) $^{63}$Cu NSLR rate 
on optimally doped YBCO.

\section{Pair propagator for unconventional pairing}
\label{pairpropagator}

\begin{figure}
\begin{minipage}{0.95\hsize}
\begin{center}
\begin{picture}(235,30)(0,-15)
\SetScale{0.5}

\SetWidth{0.5} \Line(0,-30)(50,-30) \Line(0,30)(50,30)
\GBox(5,-30)(45,30){0.9}
\SetPFont{Times-Roman}{25}
\PText(28,4)(0)[]{L}
\PText(70,5)(0)[]{=}
\SetWidth{0.5} \Line(85,15)(115,-15) \Line(85,-15)(115,15)
\SetWidth{0.5} \GCirc(100,0){10}{0.9} 
\SetPFont{Times-Roman}{15}
\PText(101,5)(0)[]{g}
\SetPFont{Times-Roman}{25}
\PText(125,5)(0)[]{+}

\SetWidth{0.5} \Line(135,15)(150,0) \Line(135,-15)(150,0)
\SetWidth{2.5} \Line(150,0)(185,-30) \Line(150,0)(185,30)
\SetWidth{0.5} \GCirc(150,0){10}{0.9} \CArc(150,0)(13,-45,45)
\SetPFont{Times-Roman}{15}
\PText(151,5)(0)[]{g}
\SetPFont{Times-Roman}{25}
\SetWidth{0.5} \Line(185,-30)(230,-30) \Line(185,30)(230,30)
\GBox(185,-30)(225,30){0.9}
\SetWidth{0.5} \Line(183,-27)(183,27)
\SetPFont{Times-Roman}{25}
\PText(208,4)(0)[]{L}

\PText(245,5)(0)[]{+}
\SetWidth{0.5} \Line(255,15)(270,0) \Line(255,-15)(270,0)
\SetWidth{2.5} \Line(270,0)(305,-30) \Line(270,0)(305,30)
\SetWidth{0.5} \GCirc(270,0){10}{0.9} \CArc(270,0)(13,-45,45)
\SetPFont{Times-Roman}{15}
\PText(271,5)(0)[]{g}
\SetPFont{Times-Roman}{25}
\SetWidth{0.5} \Line(305,-30)(350,-30) \Line(305,30)(350,30)
\GBox(305,-30)(345,30){0.9}
\SetPFont{Times-Roman}{25}
\PText(328,4)(0)[]{C}
\SetWidth{2.5} \Line(345,-30)(385,-30) \Line(345,30)(385,30)
\SetWidth{0.5} \Line(380,-30)(430,-30) \Line(380,30)(430,30)
\GBox(385,-30)(425,30){0.9}
\SetWidth{0.5} \Line(383,-30)(383,30)
\SetPFont{Times-Roman}{25}
\PText(408,4)(0)[]{L}

\end{picture}
\begin{picture}(235,40)(0,-15)
\SetScale{0.5}

\SetWidth{0.5} \Line(0,-30)(50,-30) \Line(0,30)(50,30)
\GBox(5,-30)(45,30){0.9}
\SetPFont{Times-Roman}{25}
\PText(28,4)(0)[]{C}
\PText(70,5)(0)[]{=}
\SetWidth{0.5} \Line(95,30)(105,30) \Line(95,-30)(105,-30)
\SetWidth{2.5} \DashLine(100,30)(100,-30){5}
\SetPFont{Times-Roman}{15}
\PText(101,2)(0)[]{X}
\SetPFont{Times-Roman}{25}
\PText(125,5)(0)[]{+}

\SetWidth{0.5} \Line(145,30)(155,30) \Line(145,-30)(155,-30)
\SetWidth{2.5} \DashLine(150,30)(150,-30){5}
\SetWidth{2.5} \Line(150,-30)(185,-30) \Line(150,30)(185,30)
\SetPFont{Times-Roman}{15}
\PText(151,2)(0)[]{X}
\SetPFont{Times-Roman}{25}
\SetWidth{0.5} \Line(185,-30)(230,-30) \Line(185,30)(230,30)
\GBox(185,-30)(225,30){0.9}
\SetPFont{Times-Roman}{25}
\PText(208,4)(0)[]{C}

\end{picture}
\end{center}
\end{minipage}\\[0.3cm]
\begin{minipage}{0.95\hsize}
\caption[]{
\label{cooperon}
Upper row: Diagrammatic representation of the Bethe-Salpeter
equation for the fluctuation
propagator, $L$. The vertex, $g$, represents the pairing 
interaction, the thick solid lines are  quasiparticle Green's functions, 
and the block vertex, 
$C$, represents vertex corrections due to impurity scattering.
Thin double lines symbolize vertex factors $\eta (\psi )$
due to the anisotropy of the pairing interaction.
Lower row: Diagrammatic representation of the Bethe-Salpeter equation 
for the impurity vertex corrections. 
A thick crossed line stands for the impurity scattering vertex
in the Born approximation.
The analytic forms for 
these equations are given in Eqs. (\ref{LL}) and (\ref{Ceq}).}
\end{minipage}
\end{figure}
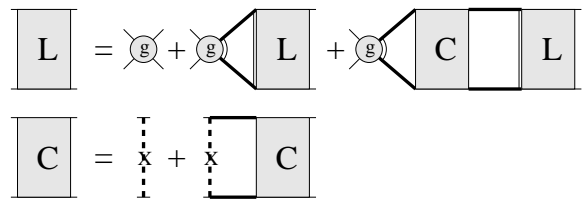

Fluctuating Cooper pairs are described by a 
propagator, $L$, which derives
from the sum over ladder diagrams in the particle-particle interaction
channel as shown in Fig. \ref{cooperon}.\cite{varlamov98}
Our derivation includes impurity scattering for a 
layered 2D superconductor with
an isotropic Fermi surface and a weak-coupling anisotropic
pairing interaction $g$. The generalization to anisotropic 2D and 3D Fermi surfaces
is straightforward.

The propagator is a function of the total momentum, $\vec{q}$,
of a pair of  interacting quasiparticles, 
their total excitation energy, $\omega$, and, for anisotropic pairing, 
their relative incoming and outgoing momenta, ${\bf k}_{in,out}$. 
In the following 
we use cylindrical coordinates ($q$, $\phi$, $q_z$) and write  ${\bf q}$ as
$
{\bf q}\ = \ \{q\cos{\phi},\ q\sin{\phi},\ q_z\}
$.
Pairing fluctuations are long-lived
only for small $\omega $ and $q$, so that the two particles which interact
have nearly opposite momenta on the Fermi surface, i.e. 
${\bf k}_{in} \approx 2{\bf k}_{F,in}$ and ${\bf k}_{out} 
\approx 2{\bf k}_{F,out}$. We
assume  a cylindrical Fermi surface of radius $k_F$, in which case the 
momenta on the Fermi surface are given by
$
{\bf k}_{F}=\{ k_F \cos{\psi},\ k_F\sin{\psi},\ k_{z}\}
$.

The pairing interaction is a function of the momenta of the
initial and final state of quasiparticles on the Fermi surface.
We denote the angles between
the $x$-axis (chosen as the tetragonal $\hat a $-axis)
and $\vec{k}_F $ and $\vec{k}_F'$ by $\psi$ and $\psi'$, respectively.
The pairing interaction, $V(\psi,\psi')$,
can be expanded in eigenfunctions belonging to the
irreducible representations of the symmetry group of the crystal.
We denote the eigenfunction  with the largest attractive (positive) 
eigenvalue by $\eta(\psi)$ and neglect for now the other sub-dominant
interactions in the expansion of $V$. 
Thus, we write the pairing interaction in the following form:
\begin{equation}
V(\psi, \psi') = 
\eta(\psi ) \cdot g \cdot \eta(\psi')\, .
\end{equation}
Note that we can neglect the small difference between
$\vec{k}_F$ and $\vec{k}_F-\vec{q}$
in the pairing interaction, since $q \sim 1/\xi_0 \ll k_F$,
where $\xi_0 = \hbar v_F/2\pi k_B T_c $ is the coherence length.

The fluctuation propagator $L(\omega, \vec{q})$
describes dynamically fluctuating Cooper pairs with a
wavelength $2\pi/q$ and a frequency $\omega$.
Near $T_c$ the typical lifetime of a pairing fluctuation 
in the clean limit is
\begin{equation}
\tau_{GL} = \frac{\hbar \pi}{8k_BT}\left(a \xi_0^2q^2+\frac{T-T_c}{T_c}
\right)^{-1},
\end{equation}
where $a= 7\zeta(3)/8 \approx 1.05$.
We set $\hbar= k_B=1$ except when explicitly noted.

In the case of strong pair breaking with dephasing time $\tau_\phi $
the prefactor $\hbar \pi /8k_BT $ is replaced by $\tau_\phi $.
Spatially small fluctuations decay faster than more extended fluctuations.
Long-lived fluctuations have typical sizes larger than 
$\xi_0 \sqrt{\frac{T_c}{T-T_c}}$.
When the temperature approaches $T_c$ the importance of long wavelength
($\vec q \to 0$),
quasi-static fluctuations ($\omega \to 0$) increases until 
fluctuation modes start to overlap in space and time.
When this happens fluctuation modes interact, which defines 
the critical fluctuation regime.
In contrast to conventional superconductors, where this regime is negligibly
small, it extends over 1-2 K in layered high-$T_c$ cuprates
like YBCO.\cite{fisher91}
Our analysis neglects interactions between fluctuation modes and
thus excludes the critical regime.

We include the effects of impurities via  the
standard  procedure of averaging over impurity positions in the limit
of a long mean free path, $\ell \gg k_F^{-1}$.\cite{Abgo}
Impurities lead to three different effects: they introduce
a finite quasiparticle lifetime  via the electron self-energy,
they generate vertex corrections, $V$, in 
the particle-hole channels, which have to be included to ensure
fundamental conservation laws, and they generate 
a Cooperon-like mode in the particle-particle channel,
the impurity vertex $C$, 
which couples directly to
the full pair fluctuation propagator, $L$.
In the case of d-wave pairing  impurities
lead to pair breaking of the pairing fluctuation modes.
We will use a short hand notation $Q\equiv (\omega_l,q,\phi)$ for
the set of arguments related to the pairing channel.
The impurity vertex (the cross in Fig. \ref{cooperon}) 
is given in terms of the impurity
scattering rate in Born approximation,
$\tilde \alpha = 1/2\pi\tau N_F$,
\begin{equation}
\label{Ceq}
C(\epsilon_n,Q)=\tilde \alpha + \tilde \alpha
A_0(\epsilon_n,Q)  C(\epsilon_n,Q) ,
\end{equation}
where $A_0(\epsilon_n,Q)$ is a momentum-averaged 
irreducible pair susceptibility, defined by the formula (with $m=0$)
\begin{eqnarray}
\label{intgf}
A_m(\epsilon_n,Q)&\equiv &A_m(\epsilon_n,\omega_l,q,\phi)=
N_F\int_0^{2\pi} d\psi \, \left[ \eta(\psi ) \right]^m
\nonumber \\
&\times &\int d\xi_{\vec k} \,
G(\epsilon_n,\xi_{\vec k}) G(\omega_l-\epsilon_n,\xi_{\vec q-\vec k}).
\end{eqnarray}
Here, $\xi_{\vec k}=\epsilon(\vec{k})-\mu $ is the quasiparticle dispersion relative
to the chemical potential.
Because $q\ll k_F$, we approximate $\xi_{\vec q-\vec k} \approx \xi_{\vec k} - \vec{v}_F\cdot \vec{q}$.
The Matsubara Green's functions are given by 
\begin{eqnarray}
G(\epsilon_n,\xi_{\vec k})&=& \frac{1}{i \epsilon_n - \xi_{\vec k}
+ \frac{i}{2} \; \mbox{sign} (\epsilon_n) 
\left( \frac{1}{\tau }+ \frac{1}{\tau_\phi} \right) },
\end{eqnarray}
where $1/\tau_\phi $ is the inelastic scattering rate and $1/\tau $ is
the elastic scattering rate.
We introduce dimensionless scattering parameters,
$\alpha = \hbar/2\pi \tau k_B T_c$
and $\alpha_\phi = \hbar/2\pi \tau_\phi k_B T_c$.
The inelastic scattering rate contributes to the quasiparticle
scattering, but not
to the impurity vertex $C$ for the fluctuation propagator.
Consequently,  the lifetime of the pair fluctuation propagator is
governed  by $\tau_\phi$. 
Note that both scattering parameters, $\alpha $ and $\alpha_\phi$,
are defined in terms of the {\it renormalized}
transition temperature $T_c\equiv T_c(\alpha,\alpha_\phi)$,
which is given by an Abrikosov-Gorkov formula\cite{Abgo,Abgo0} 
(see App. \ref{coherence}),
so that their values range from
zero (for the clean limit) to infinity (e.g. for the critical pair
breaking rate).
In high-$T_c$ cuprates the mean free path, $\ell $, is typically of the
order of 3-10 coherence lengths;\cite{tanner92} a reasonable
estimate is $\ell \approx 5\xi_0$, which corresponds to
$\alpha \approx 0.2 $.
For the pair breaking parameter (or dephasing rate),
$\alpha_\phi $, one usually assumes a much smaller value.
For example, comparison between theory and experiment 
for the $\hat c$-axis fluctuation magneto-resistance
yields 
$\tau_\phi T_c \approx 10$ in YBCO and BSCCO, corresponding to
$\alpha_\phi \approx 0.02$.\cite{dorin93}
An estimate of $\alpha_\phi $ from 
inelastic scattering of quasiparticles by phonons yields
$ \alpha_\phi\approx (\frac{k_BT}{\hbar \omega_D})^2$, which
at $T_c $ in optimally doped cuprates is 
$\approx 10^{-2}$. 
However, this weak-coupling estimate of inelastic pair breaking may be 
inappropriate if the inelastic lifetime is due to strong 
coupling to low-frequency boson modes. 
Strong coupling or large inelastic pair breaking can have
a strong effect on the pairing fluctuation corrections to the
nuclear spin lattice relaxation rate.\cite{eschrig94}
In weak coupling s-wave theory a sign change in the fluctuation
corrections to the rate occurs for 
$\alpha_\phi \approx 0.26 $.\cite{alphacrit}  
A similar sign change occurs in strong-coupling theory for a coupling
constant $\lambda \approx 2$.
Note, however, that a coupling
strength of $\lambda \approx 2 $ 
is much larger than that in conventional strong-coupling 
superconductors like lead. 
We consider
parameters $\alpha_\phi \gsim 0.26$ and $\lambda \gsim 2$ as
unreasonably large for high-$T_c$ cuprates. 
In high-$T_c$ materials pair breaking by 
inelastic scattering is probably not strong enough to produce such
qualitative changes in the behavior of the fluctuation corrections to the 
spin-lattice relaxation rate.
Thus, the remaining discussion focuses on fluctuations in
weak-coupling layered superconductors.

For a single pairing channel in an isotropic quasi 2D metal
the fluctuation propagator factorizes into
$\eta(\psi ) L(Q) \eta(\psi')$, where
$L(Q) $ obeys the Bethe-Salpeter equation:
\begin{eqnarray}
\label{LL}
&&L(Q) = 
g + T\sum_n g
A_2(\epsilon_n,Q) L(Q) \nonumber \\
&&+
T\sum_n g A_1(\epsilon_n,Q) C(\epsilon_n,Q) A_1(\epsilon_n,Q)
L(Q).
\end{eqnarray}
Inserting the Cooperon propagator, $C(\epsilon_n,Q)$, from (\ref{Ceq})
into (\ref{LL})
we can solve for $L(Q)$ in terms of the 
momentum integrated pair susceptibilities $A_m(\epsilon_n,Q)$ to obtain,
\begin{eqnarray}
 L(Q) &=& \frac{1}{g^{-1}-T\sum_n B_2(\epsilon_n,Q) },
\end{eqnarray}
where
\begin{eqnarray}
B_2(\epsilon_n,Q)&=&A_2(\epsilon_n,Q)+A_1(\epsilon_n,Q)^2 C(\epsilon_n,Q) \nonumber \\
&=&
\frac{A_2(\epsilon_n,Q) + \tilde\alpha \Big[ A_1^2
-A_0 A_2 \Big] (\epsilon_n,Q)}{1-\tilde\alpha A_0(\epsilon_n,Q)}.
\end{eqnarray}

Finally, we must include impurity vertex corrections in the
particle-particle channel to the
external vertices of the pair propagator. 
These corrections are incorporated by the replacement
$\eta L \eta' \to K$ with
\begin{eqnarray}
\label{Kpaar}
K(\epsilon_n,\epsilon_{n'},\psi , \psi' ,Q)
 & = & \tilde \eta (\epsilon_n,\psi ,Q)  \cdot  L(Q) \cdot
\tilde \eta (\epsilon_{n'},\psi' ,Q),\nonumber \\
\end{eqnarray}
where
\begin{eqnarray}
\tilde\eta(\epsilon_n,\psi ,Q) &=& \eta(\psi ) + A_1(\epsilon_n,Q)C(\epsilon_n,Q).
\end{eqnarray}
Combined with Eq. (\ref{Ceq}) this gives
\begin{eqnarray}
&&\tilde\eta(\epsilon_n,\psi ,Q) = \frac{\eta(\psi ) + \tilde\alpha \Big(  A_1(\epsilon_n,Q)-\eta (\psi ) 
A_0(\epsilon_n,Q)\Big) }{1-\tilde\alpha A_0(\epsilon_n,Q)}.\nonumber \\
\end{eqnarray}
In the case $\eta(\psi )\equiv 1$ we recover the standard vertex corrections and
pair propagator for
an isotropic s-wave superconductor,\cite{varlamov98}
\begin{eqnarray}
&&\tilde\eta_s(\epsilon_n,Q)= \frac{1}{1-\tilde\alpha A_0(\epsilon_n,Q)},\\
 &&L_{s}(Q) = \frac{1}{
\displaystyle
g^{-1}-T\sum_n \frac{A_0(\epsilon_n,Q)}{
1-\tilde\alpha A_0(\epsilon_n,Q)}}.
\end{eqnarray}

For dynamical quantities such as the
spin-lattice relaxation rate it is necessary to
analytically continue the pair propagator from
Matsubara energies to the real energy axis. This is done by
Eliashberg's technique,\cite{eliash} leading to the general result
\lrule
\begin{eqnarray}
\label{Lreal}
L(\omega,q,\phi)= \Bigg\{
N_F\ln \frac{T}{T_c} &-& \int_0^{\infty}  \frac{d\epsilon}{2\pi} \left[
\left( \tanh \frac{\epsilon-\omega/2}{2T} +
\tanh \frac{\epsilon+\omega/2}{2T} \right) \Im B_2(\epsilon,q,\phi)
-2 \tanh \frac{\epsilon }{2T} \Im B_{2c}(\epsilon) \right] + \nonumber \\
&+&i \int_0^{\infty}  \frac{d\epsilon}{2\pi}
\left( \tanh \frac{\epsilon-\omega/2}{2T} -
\tanh \frac{\epsilon+\omega/2}{2T} \right) \Re B_2(\epsilon,q,\phi)
\Bigg\}^{-1},
\end{eqnarray}
\rrule
\noindent where
\begin{eqnarray}
\label{B2}
&&B_2(\epsilon,q,\phi)=\frac{A_2(\epsilon,q,\phi) + \tilde\alpha 
\big[ A_1^2-A_2A_0\big](\epsilon,q)}{1-\tilde\alpha A_0(\epsilon,q)}, \\
&&B_{2c}(\epsilon ;T )=\frac{T_c}{T} B_2(\frac{T_c}{T}\epsilon ,\vec{q}=0;T=T_c).
\end{eqnarray}
Explicit expressions for the functions $A_0$, $A_1$ and $A_2$ are given
for s-wave and d-wave pairing in the Appendix \ref{Afunctions}.

In the long-wavelength limit it is possible to integrate Eq. (\ref{Lreal})
analytically
and express the pair fluctuation propagator for s-wave or d-wave
pairing as
\begin{eqnarray}
L_s(q,\omega)&=&N_F^{-1}\frac{1}{\epsilon_s +\xi_s^2 q^2 -i\omega \tau_s}\\
L_d(q,\omega)&=&N_F^{-1}\frac{1}{\epsilon_d +\xi_d^2 q^2 -i\omega \tau_d},
\end{eqnarray}
where the coherence lengths, $\xi_{s,d}$, static pair susceptibilities,
$\epsilon_{s,d}$ and lifetimes, $\tau_{s,d}$ are given in terms of
Digamma functions of the pair breaking
parameters (see App. \ref{coherence}).\cite{varlamov98}

We generalize the 
pairing fluctuation theory presented above
to finite magnetic fields.
We assume that the field points along the $\hat c$-axis of the crystal,
and  introduce the following dimensionless field,
\begin{equation}
b=\frac{4|e|B}{\hbar c} \left( \frac{\hbar v_F}{2\pi k_B T_c} \right)^2.
\end{equation}
The main effect of the magnetic field is to quantize the
orbital motion of the pairs. Through second order in the
momentum operator, $\vec{q}=-i\vec{\nabla}-\frac{2e}{c}\vec{A}$,
quantization of the orbital motion is achieved
by the replacements\cite{kurk72,dorin93}
\begin{eqnarray}
\label{quant}
q^2 &\to& 
\left( k+\frac{1}{2} \right) \frac{|b|}{\xi_0^2} \\
\int \frac{d^2q}{(2\pi)^2} &\to &
\quad \frac{|b|}{4\pi \xi_0^2} \sum_{k=0}^\infty
\; ,
\end{eqnarray}
where $k=0,1,\ldots $ labels the different Landau levels. 

\section{Fluctuation corrections to the nuclear spin-lattice relaxation rate}
\label{nslr}

The hyperfine interaction between quasiparticles and nuclear spins at
(fixed) lattice points $\vec{R}_\nu $
is given by
\begin{equation}
\widehat{H}_{hf}(\vec{R}_\nu ) = 
\gamma_n \gamma_e \hbar^2 \int  d^3x \, 
\widehat{\vec{I}} (\vec{R}_\nu ) \,
\underline{a}(\vec{x} -\vec{R}_\nu )\, \widehat{\vec{S}}
(\vec{x} )\, ,
\end{equation}
where $\widehat\vec{I}$ is the
nuclear spin operator, $\widehat\vec{S}(\vec{x} )$ is
the electron spin density operator, and
 $\gamma_{e,n}$ are the gyromagnetic ratios for the electron and nuclear spin,
respectively.
The coupling of the nuclei to the 
electronic system  occurs via the hermitian interaction tensor $\underline{a}\, $, which
contains the contact interaction and
dipole-dipole interaction between nuclear spin and electronic spin
density.\cite{slichter90}
The nuclear spin-lattice relaxation rate is well described by second order
perturbation theory in the hyperfine interaction between electrons  and nuclei.
The transition rate, $1/^{\nu}T^{nn^{\,\prime}}_1$, 
from nuclear state $\mid n>$ to $\mid n^{\prime}>$ 
of a nucleus at lattice point  ${\bf R}_{\nu}$ 
is determined by 
the matrix elements, $^{\nu}{\bf A}^{nn^{\prime}}_{pp^{\prime}}$, 
for the nuclear transition, accompanied
by an electronic transition from state $p$ to 
 $p^{\prime}$, and by
the imaginary part of the electronic dynamical susceptibility,
\begin{eqnarray}
&&\chi^R_{k \alpha ,p \, \beta, p' \gamma ,k' \delta }(\omega ) = \\
&&-\frac{i}{\hbar } \, \lim_{\underline{\delta }\to 0}
\int_0^{\infty } dt\,
e^{i(\omega +i\underline{\delta })t} \, \langle
[ c_{k \alpha }^+ (t)c_{p \beta }(t),
c_{p' \gamma }^+(0)c_{k' \delta }(0) ] \rangle. \nonumber
\end{eqnarray}
$c_{k \alpha }^+$ ($c_{k \alpha }$) creates (annihilates) a conduction
electron in the Bloch state labeled by $k$ with spin $\alpha $. We use
the short hand notation, $k=(\vec{k},i)$ for a Bloch state with
momentum $\vec{k}$ in band $i$.
The transition rate is given by
\begin{eqnarray}
\frac{\hbar}{\mbox{$^\nu T$}_1^{nn'}} &=&
2k_BT\sum_{k p \alpha \beta } 
\sum_{k' p'
\gamma \delta } \Big( \mbox{$^\nu\vec{A}$}_{kp }^{nn'} \gvec{\sigma }_{\alpha
\beta } \Big) \Big( \mbox{$^\nu\vec{A}$}_{p' k'}^{n'n} \gvec{\sigma }_{\gamma
\delta } \Big) \cdot \nonumber \\
&& \cdot \lim_{\omega \to 0} \frac{\Im \,
\chi^R_{k \alpha ,p \, \beta, p' \gamma ,k' \delta }
(\omega )}{\omega } \, \, .
\end{eqnarray}
The matrix elements $^{\nu}{\bf A}^{nn^{\,\prime}}_{kp}$ are smooth 
functions of the momenta. Hence, $k$ and $p$, can be
evaluated on the Fermi surface. 
In terms of Bloch wave functions 
$\phi_{k }(\vec x )$ the hyperfine matrix elements are given by
\begin{eqnarray}
\mbox{$^\nu\vec{A}$}_{kp}^{n'n}&=&
\frac{\gamma_n \gamma_e}{2} \hbar^2 \bra{n} \, \widehat\vec{I} (\vec{R}_\nu ) \ket{n'} 
\nonumber \\
&& \times
\int  d^3x \, \phi_{k }^*(\vec x ) \, \underline{a}(\vec{x} -\vec{R}_\nu )\, \,
\phi_{p }(\vec x ) \, \, .
\end{eqnarray}
and satisfy
$^{\nu}{\bf A}^{nn^{\,\prime}}_{kp}
=\left(^{\nu}{\bf A}^{n^{\,\prime}n}_{pk}\right)^{\ast}$.
In what follows we suppress 
the indices referring to the nuclear transition.

We perform a systematic expansion of $(\mbox{$^\nu T$}_1T)^{-1}$ 
in the small parameter
$T_c/E_F$ (where $E_F$ is the Fermi energy)
to obtain the leading order term of order 
$(T_c/E_F)^0$, and all corrections of order $(T_c/E_F)^1$ arising from
pair fluctuation modes. We neglect pure 
weak localization corrections
and corrections due to the temperature dependences of the hyperfine
coupling matrix elements and of the pairing interaction $g$.
Details of the classification of diagrams in terms of the small parameter
$T_c/E_F$ is given in Appendix \ref{diag}.
We evaluate the diagrams in the Appendix \ref{Corr}. 

The leading order contribution to $(\mbox{$^\nu T$}_1T)^{-1}$
is of order $(T_c/E_F)^0$
and defines the Fermi-liquid theory result for the normal state
NSLR rate,
\begin{eqnarray}
\label{korr}
&&(\mbox{$^\nu T$}_1T)^{-1}_N = 4\pi 
 \int d\vec k_F \int d\vec p_F \;  N_{\vec k_F} N_{\vec p_F} \,
| \mbox{$^\nu \vec{A}$}_{\vec k_F\vec p_F} |^2 \, ,
\end{eqnarray}
where $N_{\vec k_F}$ is the 
angle resolved quasiparticle density of states
on the Fermi surface, and $\vec{k}_F$ defines a point on the Fermi
surface.
The quasiparticle density of states is given by $N_F=\int d\vec k_F N_{\vec k_F}$.
The right hand side of Eq. (\ref{korr}) is the Korringa constant.\cite{slichter90}

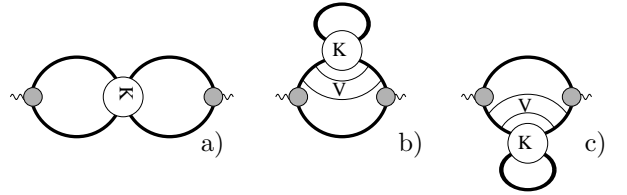
\begin{figure}
\begin{minipage}{0.95\hsize}
\begin{center}
\begin{picture}(220,40)(0,0)
\SetScale{0.5}

\SetScaledOffset(10,30)
\SetWidth{0.5} \Photon(-15,0)(0,0){2}{2.5}
\SetWidth{2.5} \CArc(35,-5)(35,5,175)
\SetWidth{2.5} \CArc(35,5)(35,-175,-5)
\SetWidth{2.5} \CArc(105,-5)(35,5,175)
\SetWidth{2.5} \CArc(105,5)(35,-175,-5)
\SetWidth{0.5} \BCirc(70,0){15}
\SetPFont{Times-Roman}{15}
\PText(75,3)(-90)[]{K}
\SetPFont{Times-Roman}{10}
\SetPFont{Times-Roman}{10}
\SetWidth{0.5} \Photon(138,0)(153,0){2}{2.5}
\SetWidth{0.5} \GCirc(2,0){7}{0.7} \GCirc(138,0){7}{0.7}
\SetScaledOffset(165,30)
\Text(74,-3)[]{a)}

\SetScaledOffset(195,30)
\SetScaledOffset(210,30)
\SetWidth{0.5} \Photon(-15,0)(0,0){2}{2.5}
\SetWidth{2.5} \CArc(35,-5)(35,5,175)
\SetWidth{2.5} \CArc(35,5)(35,-175,-5)
\SetWidth{2.5} \Oval(35,55)(15,20)(0)
\SetWidth{0.5} \BCirc(35,35){15}
\SetPFont{Times-Roman}{15}
\PText(33,38)(0)[]{K}
\SetPFont{Times-Roman}{10}
\SetPFont{Times-Roman}{10}
\SetWidth{0.5} \CArc(35,35)(23,-150,-30) \CArc(35,35)(37,-145,-35)
\SetPFont{Times-Roman}{15}
\PText(33,8)(0)[]{V}
\SetPFont{Times-Roman}{9}
\SetPFont{Times-Roman}{10}
\SetWidth{0.5} \Photon(68,0)(83,0){2}{2.5}
\SetWidth{0.5} \GCirc(2,0){7}{0.7} \GCirc(68,0){7}{0.7}
\SetScaledOffset(95,30)
\Text(149,-3)[]{b)}

\SetScaledOffset(335,30)
\SetScaledOffset(350,30)
\SetWidth{0.5} \Photon(-15,0)(0,0){2}{2.5}
\SetWidth{2.5} \CArc(35,-5)(35,5,175)
\SetWidth{2.5} \CArc(35,5)(35,-175,-5)
\SetWidth{2.5} \Oval(35,-55)(15,20)(0)
\SetWidth{0.5} \BCirc(35,-35){15}
\SetPFont{Times-Roman}{15}
\PText(33,-32)(0)[]{K}
\SetPFont{Times-Roman}{10}
\SetPFont{Times-Roman}{10}
\SetWidth{0.5} \CArc(35,-35)(23,30,150) \CArc(35,-35)(37,35,145)
\SetPFont{Times-Roman}{15}
\PText(33,-4)(0)[]{V}
\SetPFont{Times-Roman}{9}
\SetPFont{Times-Roman}{10}
\SetWidth{0.5} \Photon(68,0)(83,0){2}{2.5}
\SetWidth{0.5} \GCirc(2,0){7}{0.7} \GCirc(68,0){7}{0.7}
\SetScaledOffset(225,30)
\Text(219,-3)[]{c)}

\SetScaledOffset(520,30)

\end{picture}

\end{center}
\vspace{0.5cm}
\begin{minipage}{0.95\hsize}
\caption[]{
\label{high}
Leading order corrections in $T_c/E_F$ to the spin-lattice relaxation rate:
a) Maki-Thompson, b) and c) density of states corrections.
$V$ denotes vertex corrections in the particle-hole channel; $V=1$
in our model.
$K$ denotes the (impurity renormalized)
fluctuation mode in the pairing channel, Eq. (\ref{Kpaar}).}
\end{minipage}
\end{minipage}
\end{figure}
The fluctuation corrections to $1/T_1T$
of order $T_c/E_F$ are determined in a diagrammatic expansion
of the dynamical susceptibility
by the Maki-Thompson (MT) diagram, labeled a) in Fig. \ref{high},
and the two density of states (DOS) corrections, labeled b) and c) in Fig. \ref{high}. 
The Aslamazov-Larkin diagram (not shown) is another order smaller
in the ratio $T_c/E_F$.
The sum of these corrections can be written in the following form,
\begin{eqnarray}
\label{rate}
&&\frac{\delta (T_1T)^{-1}}{(T_1T)^{-1}_{N}} = \frac{T_c}{E_F} \times
 \nonumber \\
&&\int_0^{\infty} \frac{v_F^2 q dq}{2\pi N_F T_c } 
\int_0^{2\pi}\frac{d\phi}{2\pi}\left( S_M(q,\phi) +S_D(q,\phi) \right).
\end{eqnarray}
where the integrand is obtained by analytic continuation (following 
Eliashberg\cite{eliash}) of the Maki-Thompson and density 
of states corrections to the dynamical susceptibility obtained 
from Eqs. (\ref{wnmrMT}) and (\ref{wnmrDOS}) of Appendix \ref{Corr}, 
\begin{eqnarray}
\delta (T_1T)^{-1}=\lim_{\omega \to 0} 2\, \Im 
\frac{\chi_{MT}(\omega ) + \chi_{DOS}(\omega )}{\omega }.
\end{eqnarray}
Thus, we obtain for $S_D$ and $S_M$, 
\end{multicols}
\begin{eqnarray}
\label{S_D}
S_D(q,\phi) &=&
\frac{1}{2}\int_0^{\infty} \frac{d\omega}{2\pi} \coth \frac{\omega}{2T}
\Re L(\omega,q,\phi) \int_0^{\infty}  \frac{d\epsilon}{2\pi}
\left[ \partial^2_\epsilon \tanh \frac{\epsilon+\omega/2}{2T} -
\partial^2_\epsilon \tanh \frac{\epsilon-\omega/2}{2T} \right] \Re 
\left( B_N B_2(\epsilon,q,\phi) \right) \nonumber \\
&+&\frac{1}{2}\int_0^{\infty} \frac{d\omega}{2\pi} \coth \frac{\omega}{2T}
\Im L(\omega,q,\phi) \int_0^{\infty}  \frac{d\epsilon}{2\pi}
\left[ \partial^2_\epsilon \tanh \frac{\epsilon+\omega/2}{2T} +
\partial^2_\epsilon \tanh \frac{\epsilon-\omega/2}{2T} \right] \Im 
\left( B_NB_2(\epsilon,q,\phi) \right) \nonumber \\
&+&\int_0^{\infty} \frac{d\omega}{2\pi} 
\left( \partial_\omega \coth \frac{\omega}{2T} \right)
\Im L(\omega,q,\phi) \int_0^{\infty}  \frac{d\epsilon}{2\pi}
\left[ \partial_\epsilon \tanh \frac{\epsilon-\omega/2}{2T} -
\partial_\epsilon \tanh \frac{\epsilon+\omega/2}{2T} \right] \Im
\left( B_NB_2(\epsilon,q,\phi) \right) ,\\[0.3cm]
\label{S_M}
S_M(q,\phi) &=&
-2\int_0^{\infty} \frac{d\omega}{2\pi} \coth \frac{\omega}{2T}
\Re L(\omega,q,\phi) \int_0^{\infty}  \frac{d\epsilon}{2\pi}
\left[ \partial_\epsilon \tanh \frac{\epsilon-\omega/2}{2T} -
\partial_\epsilon \tanh \frac{\epsilon+\omega/2}{2T} \right] \Im 
\left( B_1(\epsilon,q,\phi)^2 \right) \nonumber \\
&-&2\int_0^{\infty} \frac{d\omega}{2\pi} \coth \frac{\omega}{2T}
\Im L(\omega,q,\phi) \int_0^{\infty}  \frac{d\epsilon}{2\pi}
\left[ \partial_\epsilon \tanh \frac{\epsilon-\omega/2}{2T} +
\partial_\epsilon \tanh \frac{\epsilon+\omega/2}{2T} \right] \Re 
\left( B_1(\epsilon,q,\phi)^2 \right) \nonumber \\
&+&2\int_0^{\infty} \frac{d\omega}{2\pi} 
\left( \partial_\omega \coth \frac{\omega}{2T} \right)
\Im L(\omega,q,\phi) \int_0^{\infty}  \frac{d\epsilon}{2\pi}
\left[ \tanh \frac{\epsilon-\omega/2}{2T} -
\tanh \frac{\epsilon+\omega/2}{2T} \right]
\left| B_1(\epsilon,q,\phi) \right|^2,
\end{eqnarray}
\rrule
\noindent where $B_N=2\pi N_F$, $B_2(\epsilon,q,\phi)$ is
defined in (\ref{B2}) and
\begin{eqnarray}
&& B_1(\epsilon,q,\phi)=\frac{A_1(\epsilon,q,\phi)}{1-\tilde\alpha A_0(\epsilon,q)}\, .
\end{eqnarray}
The Fermi energy, $E_F$, is related to measurable properties
of the 2D Fermi liquid by $E_F=\hbar^2 v_F^2 \pi N_F a_c$,
where $a_c$ is the $\hat c$-axis dimension of the unit cell.
Equation (\ref{S_D}) originates from corrections to the rate
due to pairing fluctuation corrections to the quasiparticle density of states,
Fig. \ref{high} b) and c).
The first two terms in Eq. (\ref{S_D}) also determine the fluctuation corrections
to the Pauli spin susceptibility, which we discuss in section \ref{Susc}.
Equation (\ref{S_M}) represents the Maki-Thompson corrections.
The first two terms in (\ref{S_M}) are referred
to as the ``regular'' Maki-Thompson contribution,
and the last term is the ``anomalous'' Maki-Thompson
contribution.
The regular MT contribution gives a negative correction as does the DOS term.
The anomalous MT term is positive, but its magnitude is
very sensitive to pair breaking processes.
This is the basis for differentiating s-wave and d-wave pairing
fluctuations using NMR.

Results for the fluctuation corrections in the
quasi-static limit are obtained by expanding the integrand
for small $\omega $ (the singularities
of the coth-factors are removable). The long-wavelength limit follows by
expanding the denominator of the pair propagator
to second order in $q$ and approximating the remaining terms in the 
integrals by their limits for 
$q\to 0$. Results for $\delta (1/T_1T) $ in these
limits are discussed by Randeria and Varlamov.\cite{randeria94}
We did not make these approximations;
rather we performed the $\phi$-integral analytically and the integrals over
$\epsilon $ and $\omega $ numerically.
As we discuss later in this section, our approach is important for extending
the theory such that a quantitative comparison with high-field NMR experiments
can be made.

In a magnetic field with $H||\hat c$, the orbital motion of the
pairing fluctuations is quantized.
Landau level quantization is achieved by the
replacements shown in Eq. (\ref{quant}).
Fluctuation corrections in a magnetic field are often
treated in the small field limit, where an expansion 
in the magnetic field up to second order is performed.
At high fields a common approximation is to retain only the
lowest Landau level.
However, one is often in the regime between these limits. 
This is the case for the recent high-field NMR experiments in 
YBCO.\cite{carretta96,mitrovic98,bachman98}
To analyze this regime we sum the 
fluctuations over the Landau levels numerically.
We introduce a cut-off field,
$b_c$, to regulate the sum over Landau levels, which
would otherwise lead to logarithmically
divergent fluctuation corrections.
This divergence is an artifact of the (standard) approximation
$\xi_{\vec q-\vec k}\approx \xi_{\vec k} - \vec{v}_F \cdot \vec{q} $ made in evaluating (\ref{intgf}). 
Without this approximation convergence is achieved on a large momentum scale,
$\sim k_F$, or correspondingly for $n$ large compared to $1/b$.
We simulate the convergence for large $n$ by a cut-off field
$b_c=20$ in our numerical calculations. 
Thus, the sum over the Landau levels in (\ref{quant}) extends up to $b_c/b$.
Changes in $b_c$
lead only to overall shifts of the results, indicating small field- and 
temperature-independent ``high-energy'' corrections.
These high-energy terms renormalize the leading order relaxation
rate as discussed below.

\subsection{Results: Magnetic field dependence}
\label{magnmr}

\begin{figure}
\begin{minipage}[t]{0.48\hsize}
\centerline{\epsfxsize0.93\hsize\epsffile{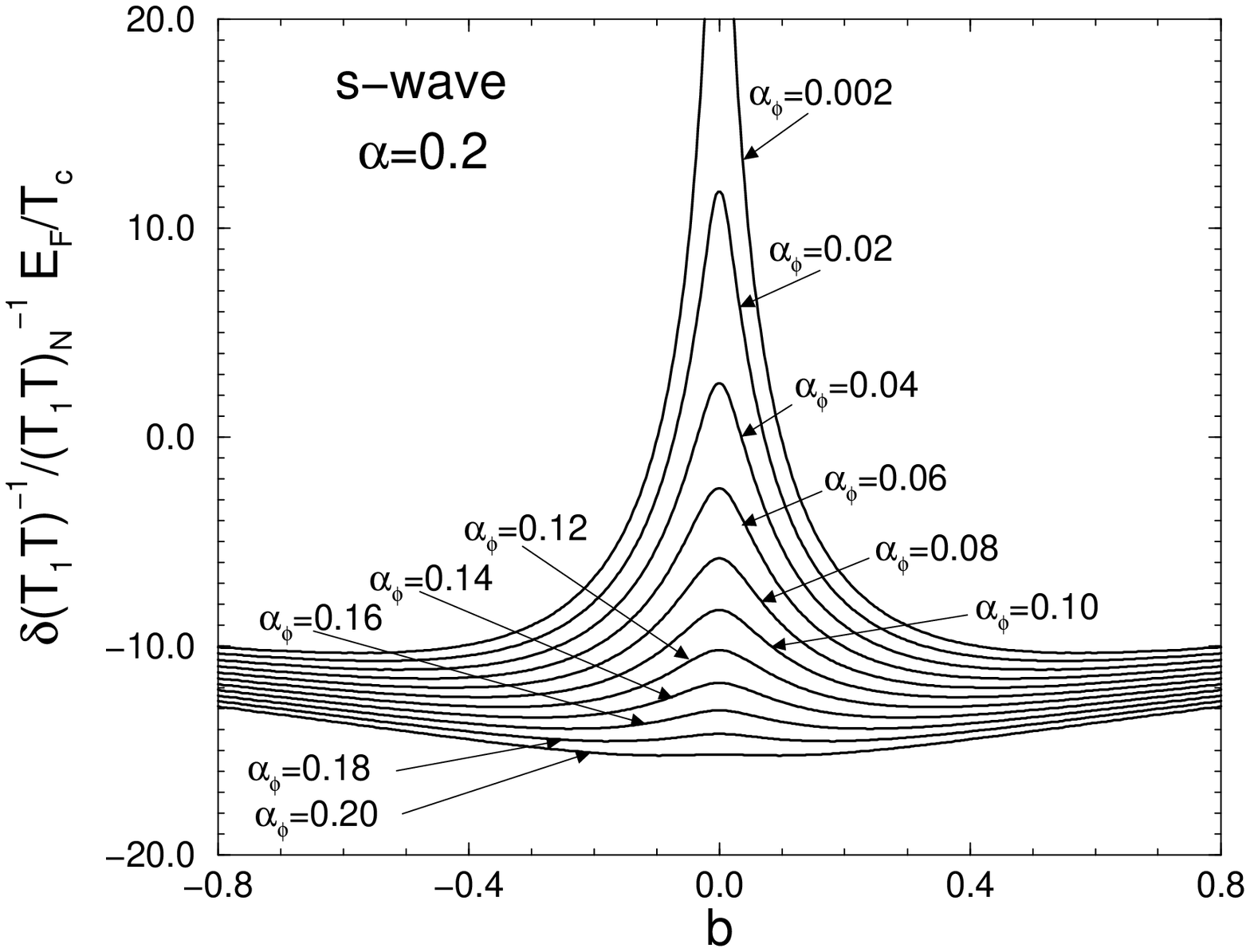}}
\begin{minipage}{0.95\hsize}
\caption[]{
\label{F3s}
Corrections to the spin-lattice relaxation rate for $T/T_c=95\mbox{K}/92.5\mbox{K} \approx 1.03$ from
s-wave pairing fluctuations
as a function of the reduced magnetic field, $b$. The elastic scattering parameter
is $\alpha=0.2$, and the pair breaking parameter, $\alpha_\phi $ varies
as indicated.}
\end{minipage}
\end{minipage}
\begin{minipage}[t]{0.48\hsize}
\centerline{\epsfxsize0.93\hsize\epsffile{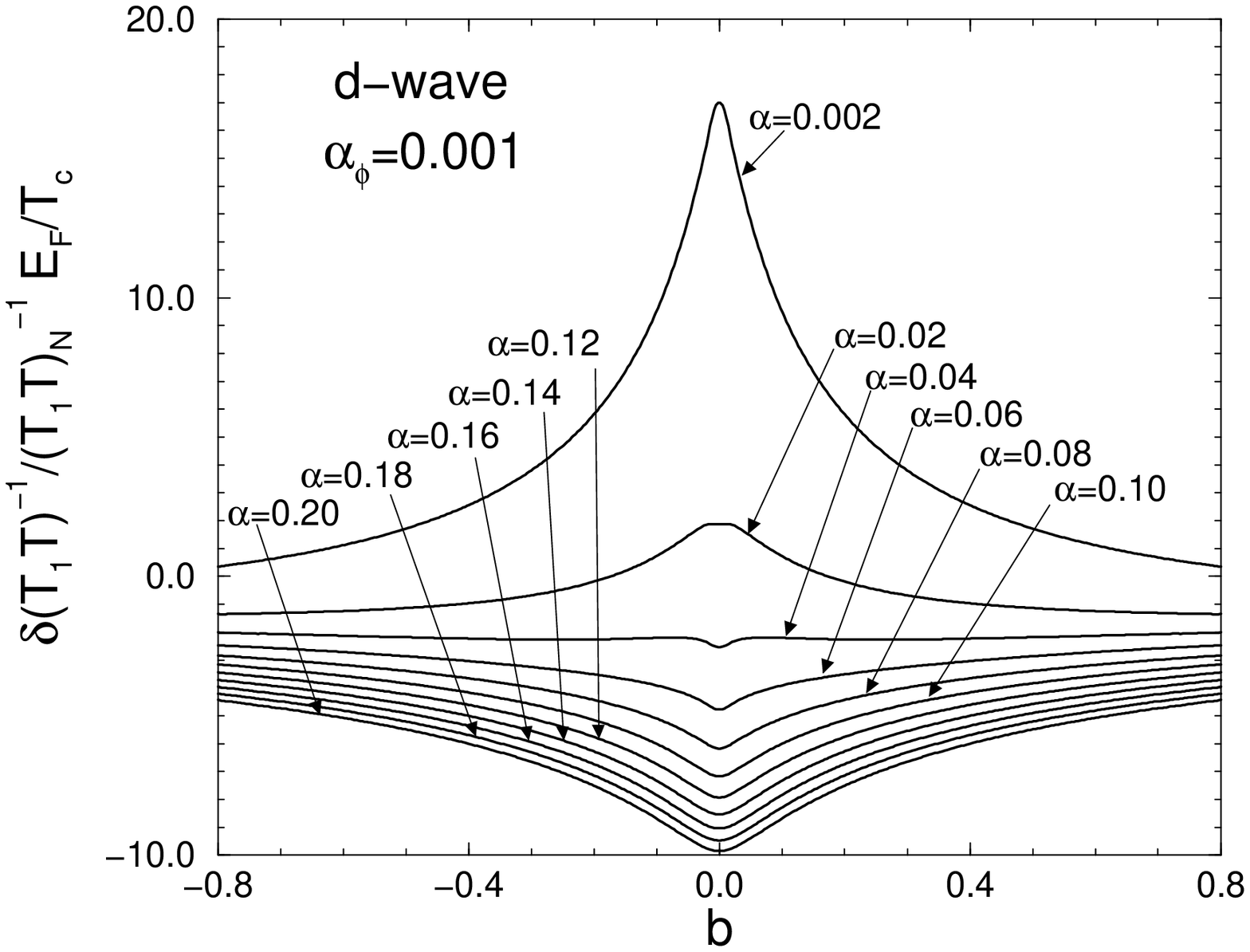}}
\begin{minipage}{0.95\hsize}
\caption{
\label{F3d}
Corrections to the spin-lattice relaxation rate for $T/T_c=95\mbox{K}/92.5\mbox{K} \approx 1.03$ from
d-wave pairing fluctuations
as a function of the reduced magnetic field, $b$.
The pair breaking parameter is $\alpha_\phi=0.001$, and the elastic 
scattering parameter, $\alpha  $, varies as indicated.}
\end{minipage}
\end{minipage}
\end{figure}

Calculations of the fluctuation corrections
to the NSLR rate are shown in Fig. \ref{F3s} (for s-wave pairing)
and Fig. \ref{F3d} (for d-wave pairing). We normalized the
results by dividing out the small prefactor, $(T_c/E_f)$, and
the normal-state NSLR rate, $(T_1T)^{-1}_N$; thus, we plot the
dimensionless quantity $[\delta(T_1T)^{-1}/(T_1T)^{-1}_N]\cdot (E_F/T_c)$.
Pairing fluctuation corrections in two dimensions
contain contributions that are constant in temperature and magnetic field.
The exact values of these constants are weakly dependent 
on the cut-off in the Landau-level summation as 
mentioned above.
These constants, which appear as 

\end{multicols}
\begin{figure}
\begin{minipage}[t]{0.48\hsize}
\centerline{\epsfxsize0.93\hsize\epsffile{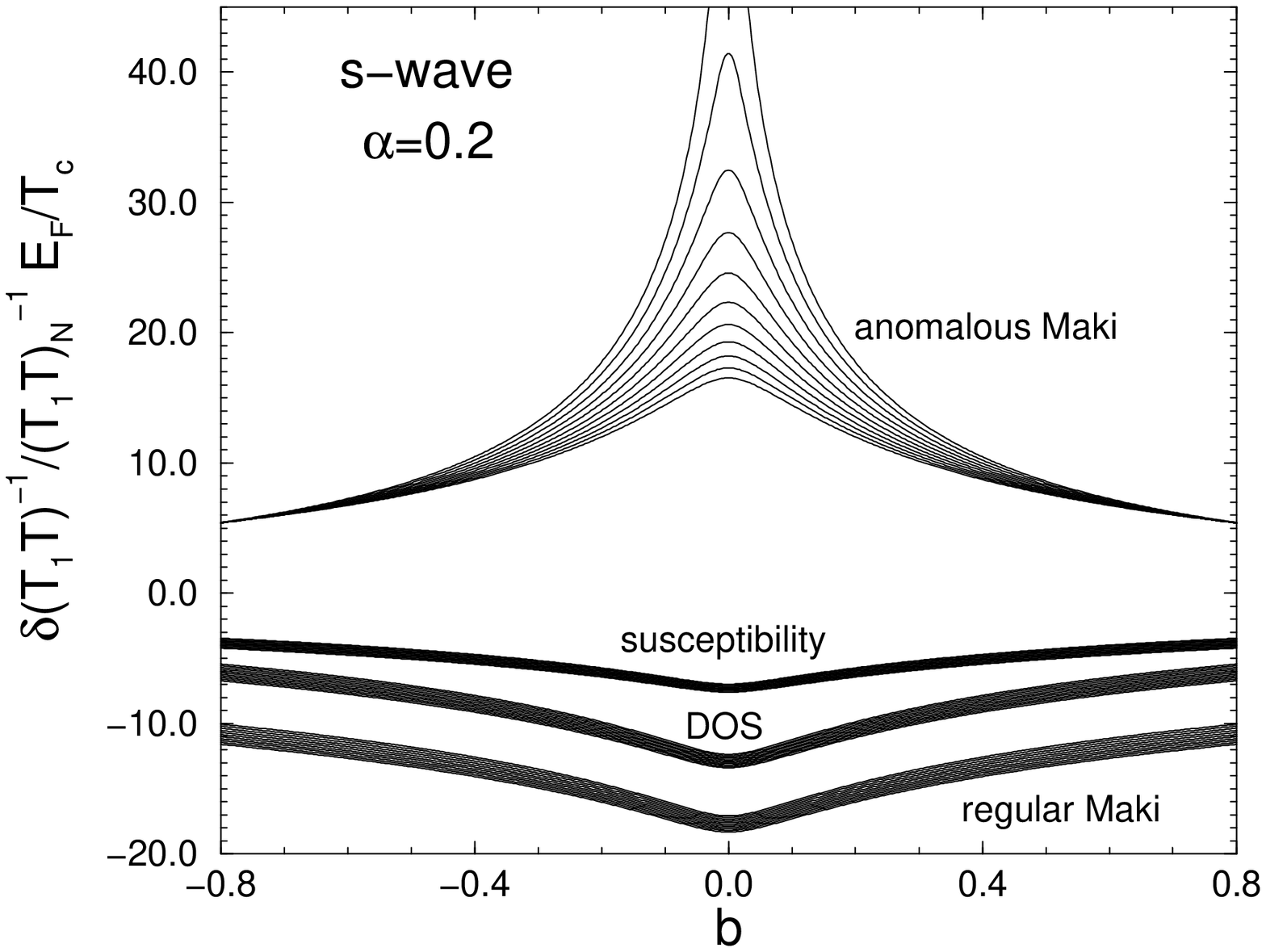}}
\begin{minipage}{0.95\hsize}
\caption[]{
\label{F1s}
Magnetic field dependence of the Maki-\-Thomp\-son
contributions and 
the DOS contribution to the fluctuation corrections to the NSLR rate
for $T/T_c=95$K$/92.5$K $\approx 1.03$, assuming s-wave pairing.
For comparison, we also show the fluctuation corrections to the 
Pauli spin susceptibility. The curves correspond to
$\alpha_\phi = 0.002$ and 0.02-0.2 (in steps of 0.02)
from top to bottom for each set. }
\end{minipage}
\end{minipage}
\begin{minipage}[t]{0.48\hsize}
\centerline{\epsfxsize0.93\hsize\epsffile{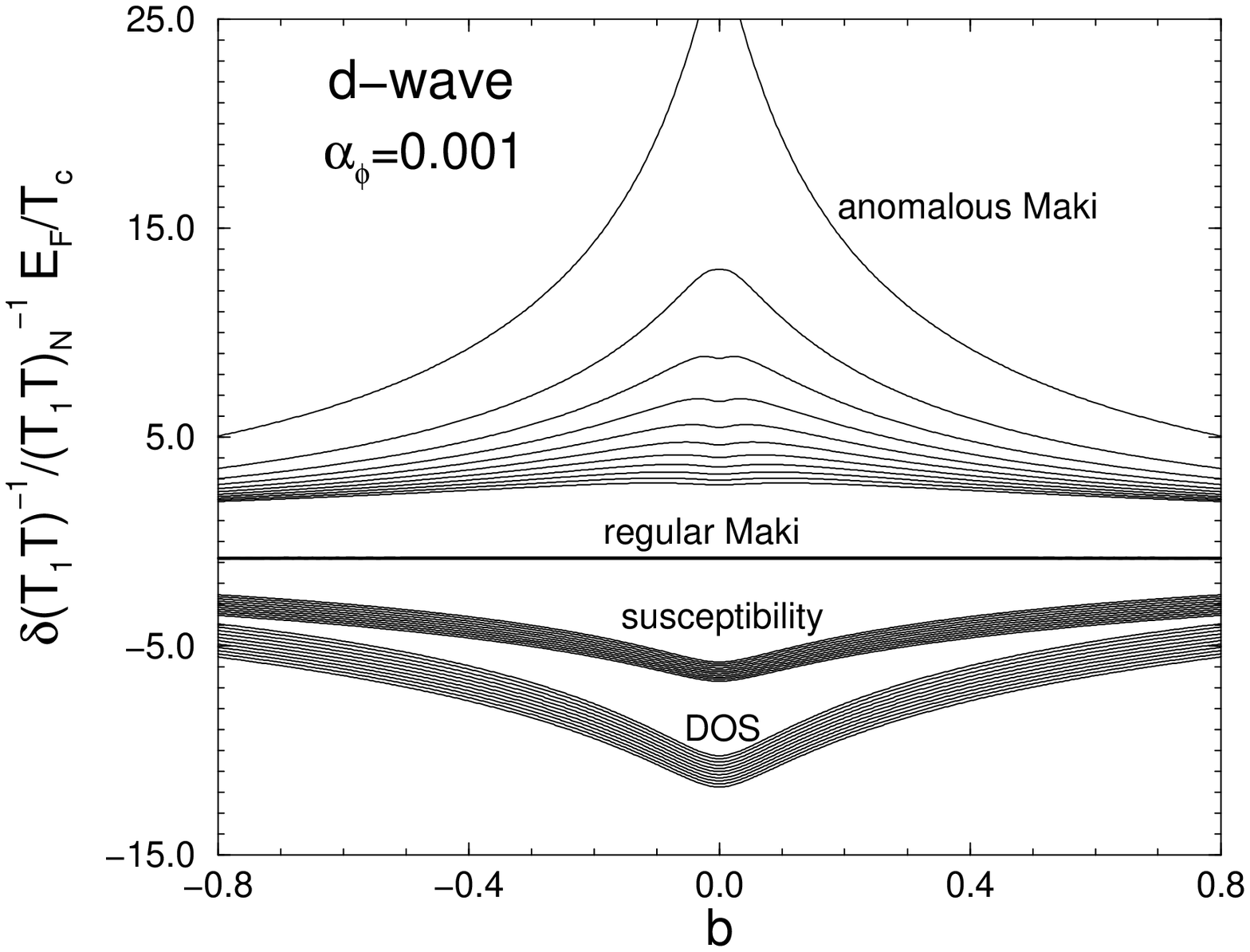}}
\begin{minipage}{0.95\hsize}
\caption[]{
\label{F1d}
The same as in Fig. \ref{F1s}, but for d-wave pairing;
$\alpha = 0.002$ and 0.02-0.2 (in steps of 0.02)
from top to bottom for each set.
Note, that the anomalous Maki-Thompson term dominates for
very clean systems, $\alpha \lsim 0.04$, but is negligible
for $\alpha \gsim 0.1$.}
\end{minipage}
\end{minipage}
\end{figure}
\begin{multicols}{2}
\noindent
offsets of the curves in
all following figures, are irrelevant 
and simply renormalize the normal-state rate, $(T_1T)^{-1}_N$.

Our calculations for the fluctuation corrections to $1/T_1T$ for
s-wave and d-wave pairing symmetry include pairbreaking processes
from elastic electron-impurity scattering and inelastic scattering
by emission and absorption of phonons. For s-wave symmetry we
fixed the elastic scattering rate at $\alpha=0.2$, and plotted
the corrections for the pair breaking parameter, $\alpha_\phi$,
ranging from 0.002 to 0.2. However, for d-wave symmetry non-magnetic
impurities are already pair breaking, so we fixed $\alpha_\phi =0.001$
(this value affects the results only in the ultra-clean case)
and calculated the fluctuation corrections for impurity scattering
rates ranging from $\alpha =0.002$ to $0.2$. Our results are shown
in Figs. \ref{F3s} and \ref{F3d}.
Note that the lowest curve in the d-wave case (Fig. \ref{F3d}) and the highest
curve in the s-wave case (Fig. \ref{F3s}) correspond to similar impurity and
inelastic scattering rates, and that s-wave and d-wave pairing
fluctuations show the opposite field evolution in the limit
$\alpha_{\phi}\ll\alpha\simeq 0.2$.
Furthermore, the s-wave fluctuation corrections to the NSLR rate
decrease with increasing field even for inelastic rates as large
as 
\end{multicols}
\begin{figure}
\begin{minipage}[t]{0.48\hsize}
\centerline{\epsfxsize0.93\hsize\epsffile{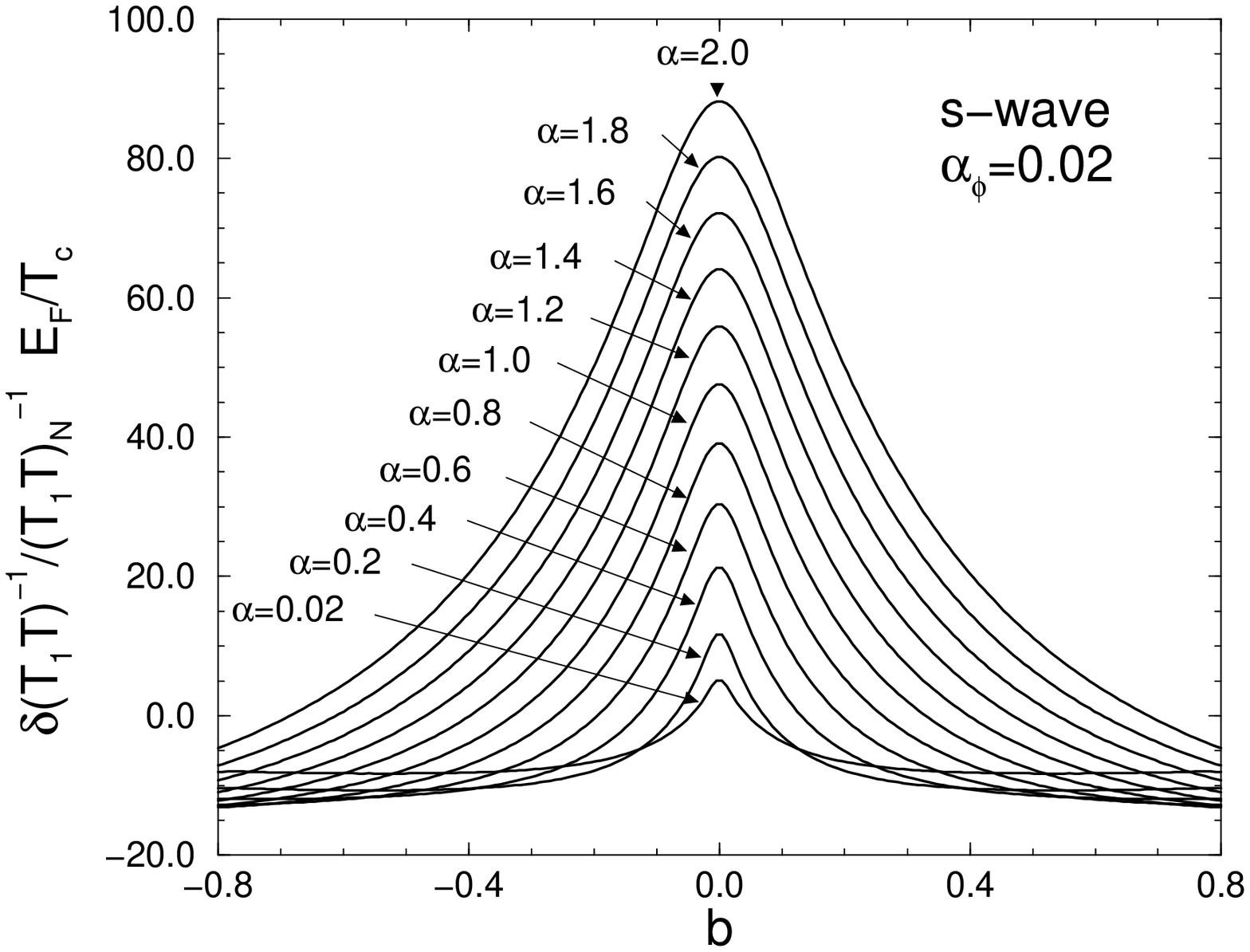}}
\begin{minipage}{0.95\hsize}
\caption[]{
\label{F3sbig}
Fluctuation corrections to NSLR rate for $T/T_c=95$K$/92.5$K $\approx 1.03$,
for $\alpha $ ranging from the clean to the dirty limit.}
\end{minipage}
\end{minipage}
\begin{minipage}[t]{0.48\hsize}
\centerline{\epsfxsize0.93\hsize\epsffile{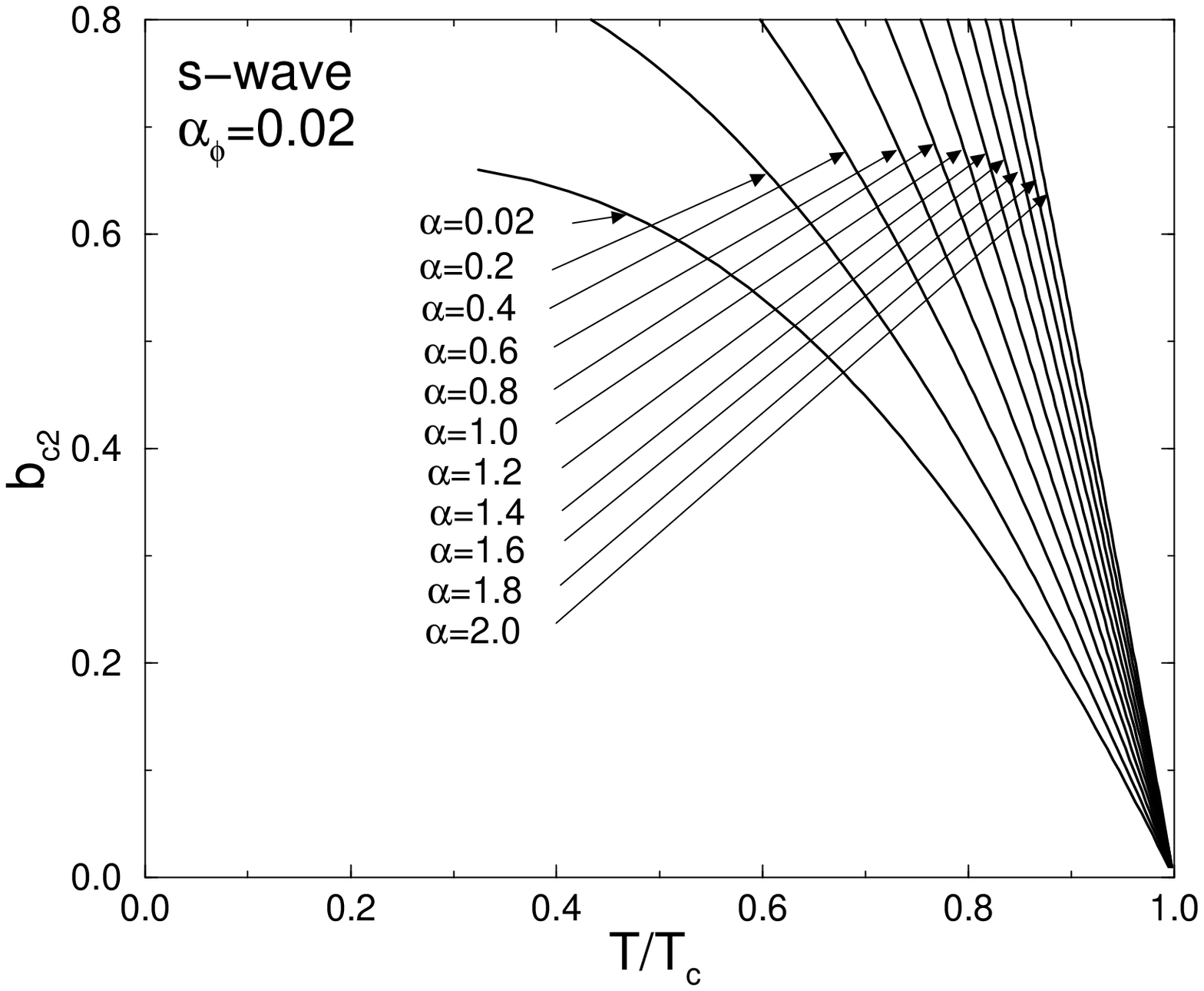}}
\begin{minipage}{0.95\hsize}
\caption[]{
\label{hc2}
Upper critical field, $b_{c2}$, for $\alpha $ ranging from the 
clean to the dirty limit.}
\end{minipage}
\end{minipage}
\end{figure}
\begin{multicols}{2}
\noindent
$\alpha_{\phi}\simeq 0.1$. For very large inelastic rates, $\alpha_{\phi}\agt 0.2$, the 
maximum in $\delta(T_1T)^{-1}$ at $b=0$ is suppressed.
Such a large inelastic pairbreaking parameter appears unlikely
for the cuprates. More realistic estimates for the elastic and inelastic
pairbreaking parameters are $\alpha_{\phi}\simeq 0.02$ and
$\alpha\simeq 0.2$.\cite{tanner92,dorin93}

For d-wave pairing the fluctuation correction to the NSLR rate
changes sign for $\alpha \approx 0.03$; the rate 
decreases with increasing field in the ultra-clean limit and
increases with increasing field in the limit of weak disorder,
$\alpha > 0.03$. In Fig. \ref{F3d} we note the rapid drop
in the rate with increasing field in the ultra-clean limit
($\alpha=0.002$) compared with the increase in the rate
with increasing field shown for $\alpha =0.2$.
It is worth noting that this behavior is not obtained
in the long-wavelength approximation employed by other
authors.\cite{varlamov98} 
We also note that in the clean limit for d-wave pairing the long-wavelength
approximation is not justified for $\frac{T-T_c}{T_c}\gsim (\alpha + \alpha_\phi)$.\cite{anomMT}

In  Figs. \ref{F1s} and \ref{F1d} we show the different contributions
to the relaxation rate for s-wave and d-wave symmetries.
The label `DOS' refers to the density of states corrections in
Eq. (\ref{S_D}). The `regular Maki' contribution
is the first two terms in Eq. (\ref{S_M}) and the `anomalous Maki'
correction corresponds to the last term in Eq. (\ref{S_M}).
The full fluctuation correction to $\delta(T_1T)^{-1}$,
shown in Figs. \ref{F3s} and \ref{F3d}, is the sum of the `DOS', 
`regular Maki', and `anomalous Maki' corrections.  The `DOS'
term also determines the fluctuation correction to the tunneling
density of states at zero bias for an NIS tunnel junction.
The fluctuation corrections to the spin susceptibility are also
shown for comparison in Figs. \ref{F1s} and \ref{F1d}.

For s-wave pairing the regular Maki-Thompson correction
is (up to a constant) nearly equal to the DOS contribution. 
By contrast, the regular Maki-Thompson term
is negligible for d-wave pairing.
All fluctuation corrections except the anomalous Maki-Thompson term
are weakly dependent on the scattering parameters in the range of interest.
The anomalous Maki-Thompson correction is extremely
sensitive to pair breaking, as can be seen in Figs. \ref{F1s} and 
\ref{F1d}. Because pair breaking by disorder is sensitive to the
symmetry of the pairing fluctuations,
the relative correction
to the NSLR rate, shown in Figs. \ref{F3s} and \ref{F3d},
shows qualitatively different behavior for s-wave and d-wave pairing
symmetries. 

In Fig. \ref{F3sbig} we show the influence of strong disorder on
the magnetic field dependence of the NSLR rate for an s-wave
superconductor. Disorder leads to a reduction of the coherence
length, and thus to an enhancement of fluctuations.
In the clean limit the typical magnitude of the fluctuation corrections
in 2D contains the prefactor, $T_c/E_F$, which is replaced
in the dirty limit ($\alpha=1/2\pi\tau T_c\gg 1$) by 
$1/\tau E_F \sim \alpha T_c/E_F$, which means that the fluctuations 
in dirty s-wave superconductors are typically stronger than fluctuations
in clean s-wave superconductors with the same $T_c$. By comparison,
d-wave superconductivity is completely suppressed by elastic scattering
for $1/2\pi \tau T_{c0} \agt 0.28$, where $T_{c0}$ is the transition
temperature without impurities.

Note that the NSLR rate for s-wave pairing decreases with increasing
field in both the clean and dirty limit for realistic
pair breaking parameters $\alpha_\phi \lsim 0.2$.
The enhancement of fluctuation corrections to the
rate reflects the
reduction in the coherence length by elastic and inelastic scattering.
For weak impurity scattering 
the reduction of the coherence length at $T_c$ for s-wave pairing
becomes
\begin{eqnarray}
\frac{\xi_s(T_c)^2}{\xi_0(T_c)^2}&=& 
a - \frac{\pi^4}{32} (\alpha_\phi + \frac{\alpha}{3}),
\end{eqnarray}
\end{multicols}
\begin{figure}
\begin{minipage}[t]{0.48\hsize}
\centerline{\epsfxsize0.93\hsize\epsffile{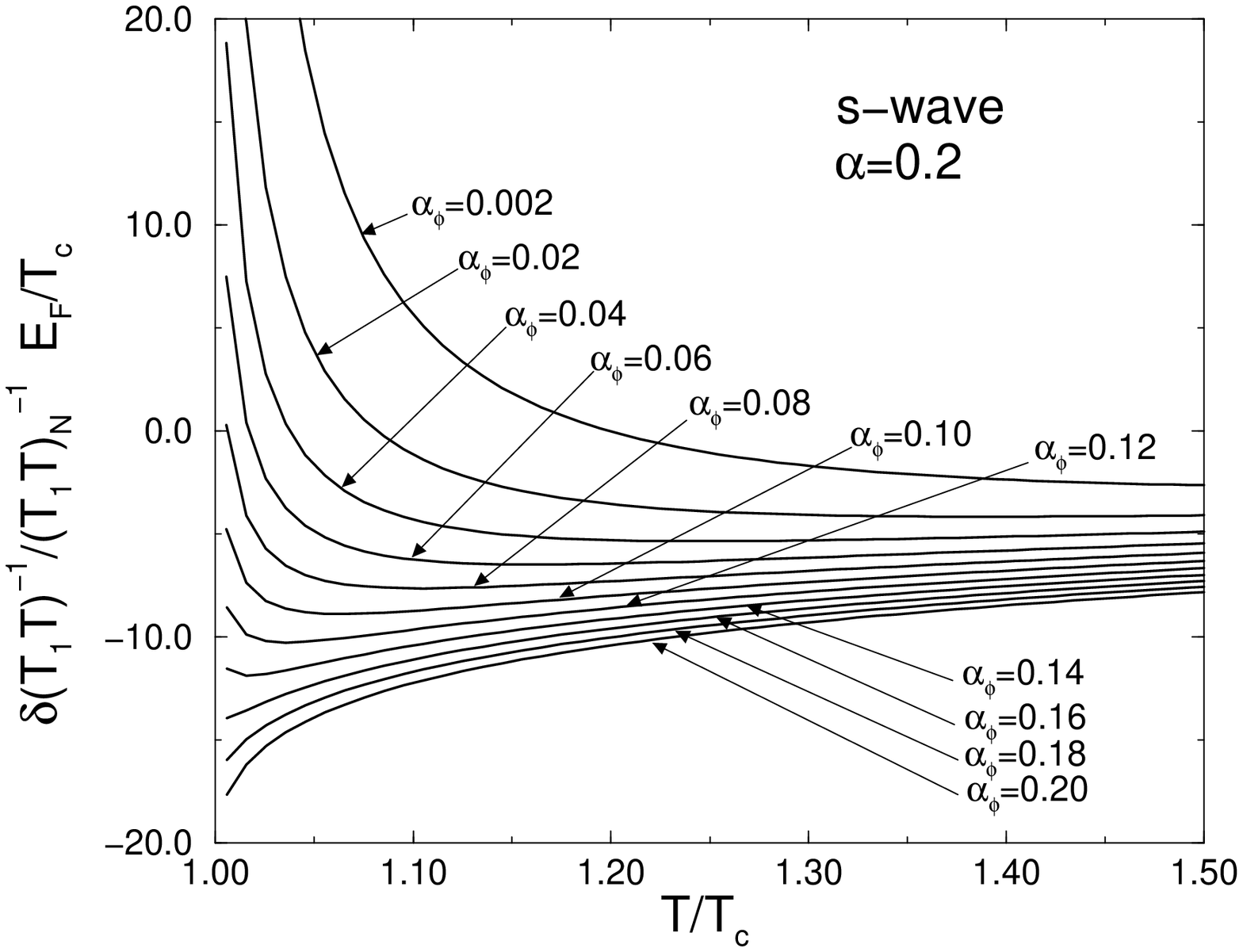}}
\begin{minipage}{0.95\hsize}
\caption[]{
\label{T1s}
Temperature dependence of
fluctuation corrections to NSLR rate for $b=0.01$, $\alpha=0.2$
 and s-wave pairing.}
\end{minipage}
\end{minipage}
\begin{minipage}[t]{0.48\hsize}
\centerline{\epsfxsize0.93\hsize\epsffile{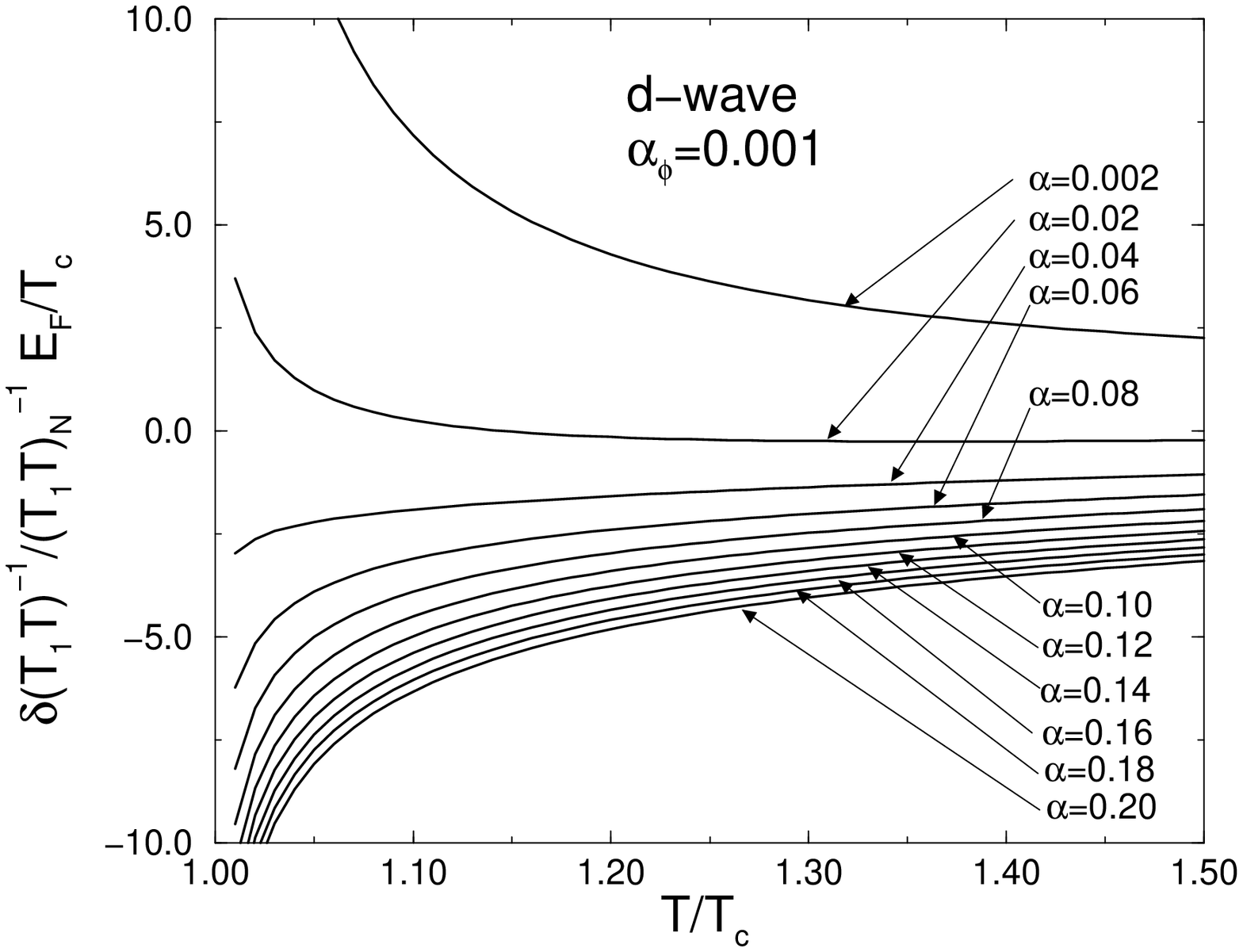}}
\begin{minipage}{0.95\hsize}
\caption[]{
\label{T1d}
Temperature dependence of
fluctuation corrections to NSLR rate for $b=0.01$, $\alpha_\phi=0.001$
 and d-wave pairing.}
\end{minipage}
\end{minipage}
\end{figure}
\begin{multicols}{2}
\noindent
and for d-wave pairing
\begin{eqnarray}
\frac{\xi_d(T_c)^2}{\xi_0(T_c)^2}&=& 
a - \frac{\pi^4}{32} (\alpha_\phi + \alpha) ,
\end{eqnarray}
where $a= 7\zeta(3)/8 \approx 1.05$.
Thus, the reduction of the
coherence length by non-magnetic impurities is stronger
by a factor of 3 for d-wave pairing compared to s-wave 
pairing at the same $T_c$.
This shortening of the coherence length
is accompanied by a suppression of the transition
from $T_{c0}=T_c(1+\frac{\pi^2}{4}(\alpha_\phi +\alpha))$ to $T_c$ 
in d-wave symmetry, compared to $T_{c0}=T_c (1+\frac{\pi^2}{4}\alpha_\phi )$
for s-wave pairing.\cite{Abgo,Abgo0}

The slope of $b_{c2}$ at $T_c$ is inversely proportional to the
square of the coherence length,
\begin{eqnarray}
\label{slope}
\frac{db_{c2,s(d)}}{dT}\bigg|_{T=T_c}=-2\; \frac{\xi_0^2}{\xi_{s(d)}^2} \; \frac{d\epsilon_{s(d)}}{dT}\bigg|_{T=T_c}.
\end{eqnarray}
Thus, the reduction in the coherence length leads to a significant increase
in the slope of $b_{c2}(T)$ shown in Fig. \ref{hc2}. 
These results were obtained by numerically solving the equation
\begin{eqnarray}
\label{hc2eq}
&&\ln \frac{T}{T_c} = N_F^{-1}\int_0^{\infty}  \frac{d\epsilon}{2\pi} 
2\tanh \frac{\epsilon}{2T} 
\nonumber \\ &&\times 
\Im \left[ B_2(\epsilon,q_b;T) -B_{2c}(\epsilon;T) \right] ,
\end{eqnarray}
where $q_b=\sqrt{|b_{c2}(T)|/2\xi_0^2}$.\cite{LLQ} 

For weak impurity scattering we obtain
\begin{eqnarray}
\frac{db_{c2}}{dT}\bigg|_{T_c}&=& 
-\frac{2}{a T_c} \left(1-\frac{\pi^2\alpha_\phi }{4} + 
\frac{\pi^4}{32a}\left( \alpha_\phi+\frac{\alpha}{3}\right) \right) \\
&\approx& -\frac{1}{T_c} (1.90 + 0.81 \alpha_\phi + 1.83 \alpha ),
\end{eqnarray}
for s-wave pairing, and 
\begin{eqnarray}
\frac{db_{c2}}{dT}\bigg|_{T_c}&=& 
-\frac{2}{a T_c} \left(1-\frac{\pi^2(\alpha+\alpha_\phi) }{4} + 
\frac{\pi^4}{32 a }\left( \alpha_\phi+\alpha\right)  \right) \\
&\approx& -\frac{1}{T_c} (1.90 + 0.81 \alpha_\phi + 0.81 \alpha ),
\end{eqnarray}
for d-wave pairing.
The negative terms in the brackets come from the reduction
of the transition temperature by pair breaking.

\subsection{Results: Temperature dependence}

\begin{figure}
\begin{minipage}[t]{0.48\hsize}
\centerline{\epsfxsize0.93\hsize\epsffile{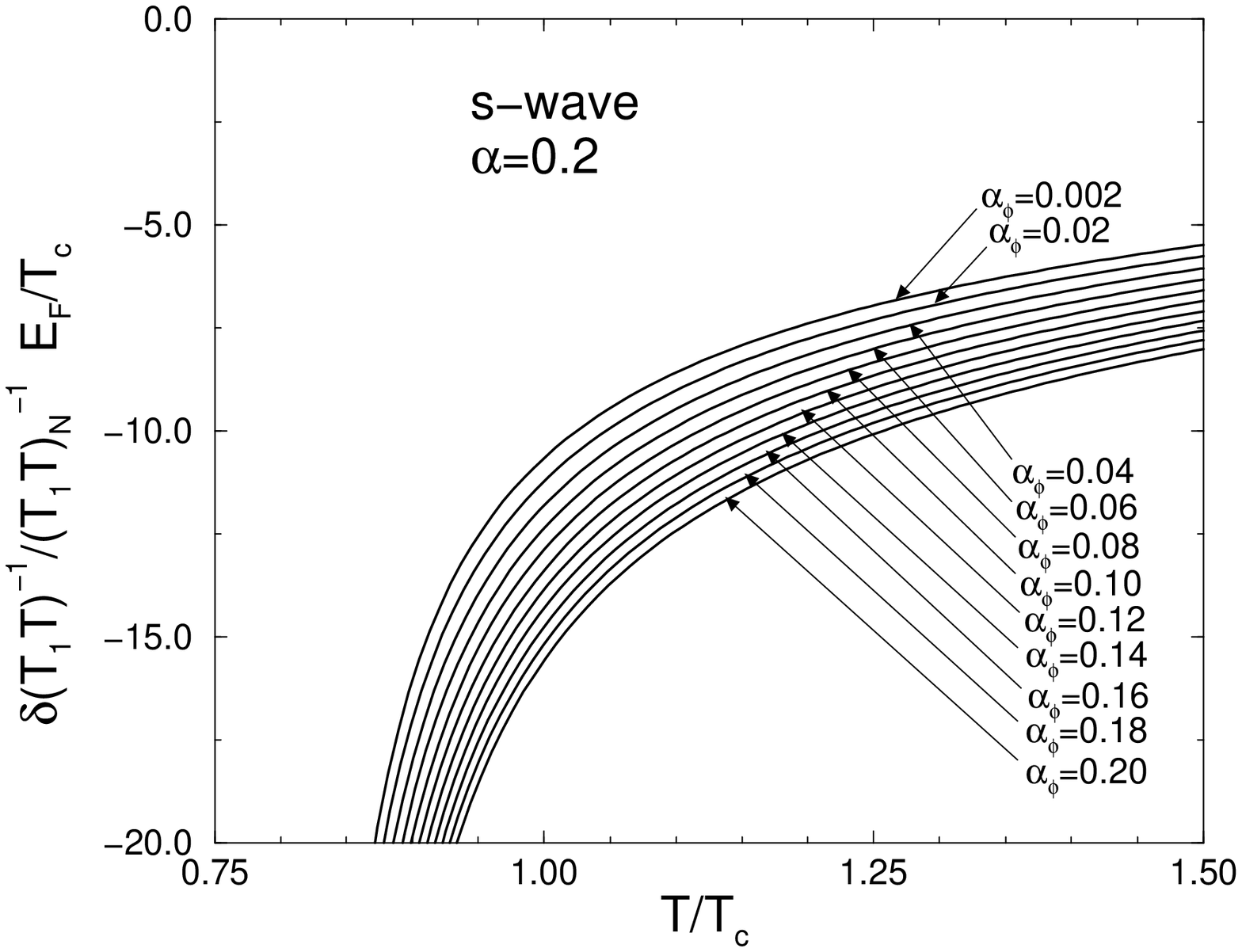}}
\begin{minipage}{0.95\hsize}
\caption{
\label{T2s}
Temperature dependence of
fluctuation corrections to NSLR rate for $b=0.4$, $\alpha=0.2$ 
and s-wave pairing.}
\end{minipage}
\end{minipage}
\begin{minipage}[t]{0.48\hsize}
\centerline{\epsfxsize0.93\hsize\epsffile{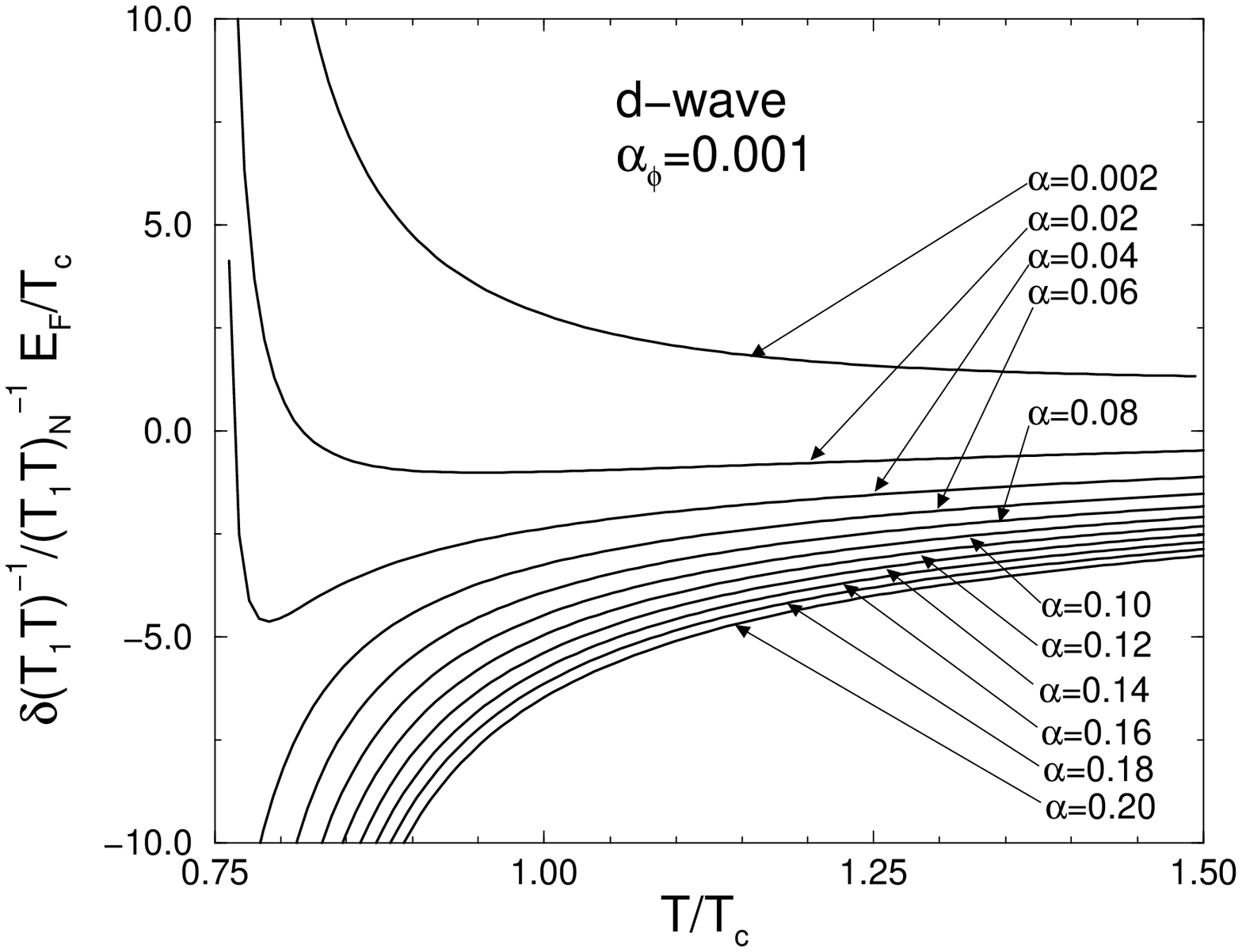}}
\begin{minipage}{0.95\hsize}
\caption{
\label{T2d}
Temperature dependence of
fluctuation corrections to NSLR rate for $b=0.4$, $\alpha_\phi=0.001$
and d-wave pairing.}
\end{minipage}
\end{minipage}
\end{figure}

\begin{figure}
\begin{minipage}[t]{0.48\hsize}
\centerline{\epsfxsize0.93\hsize\epsffile{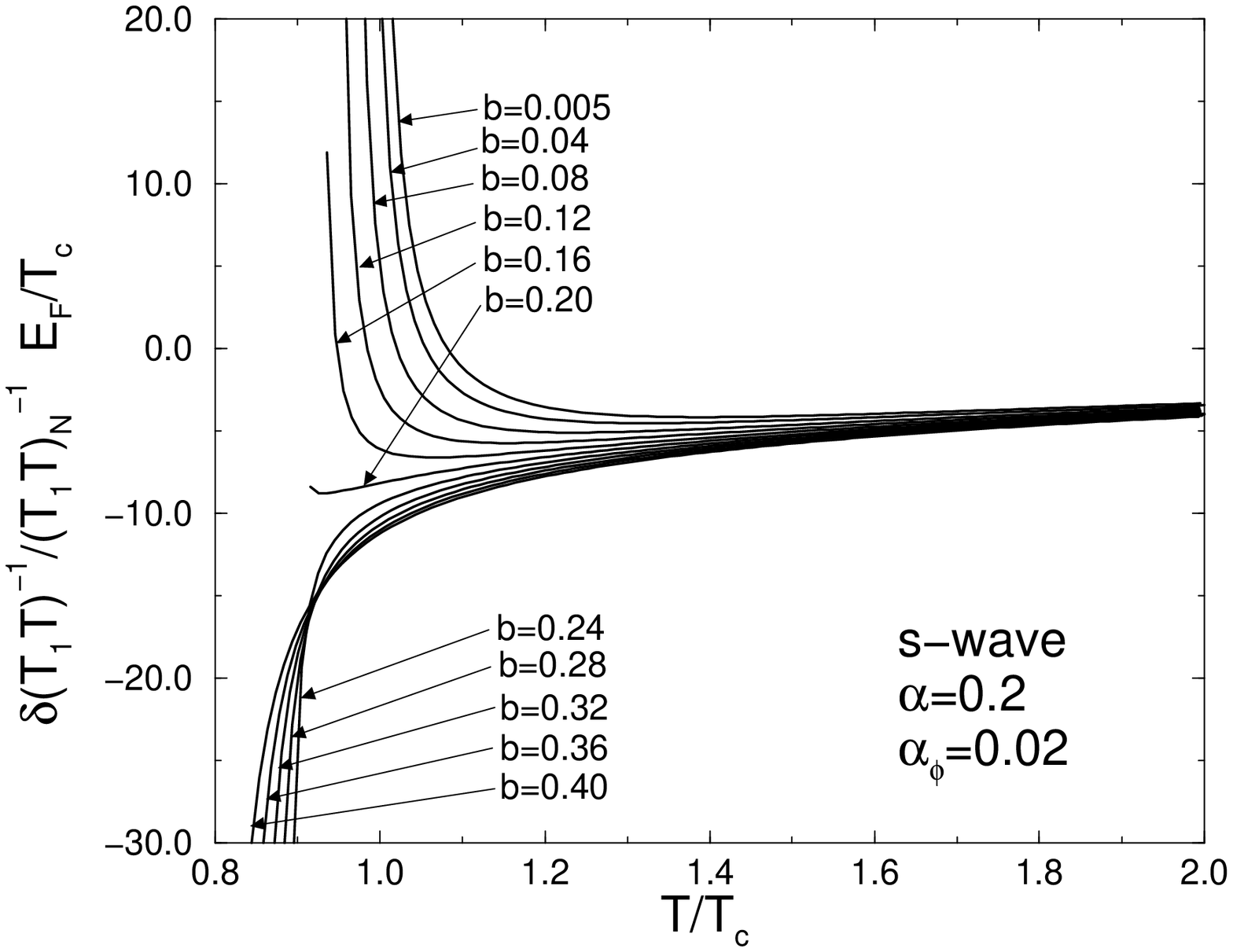}}
\begin{minipage}{0.95\hsize}
\caption{
\label{T1soverb2}
Temperature dependence of s-wave
fluctuation corrections to NSLR rate for different fields, given
as the sum of `anomalous Maki', `regular Maki' and `DOS' terms.}
\end{minipage}
\end{minipage}
\begin{minipage}[t]{0.48\hsize}
\centerline{\epsfxsize0.93\hsize\epsffile{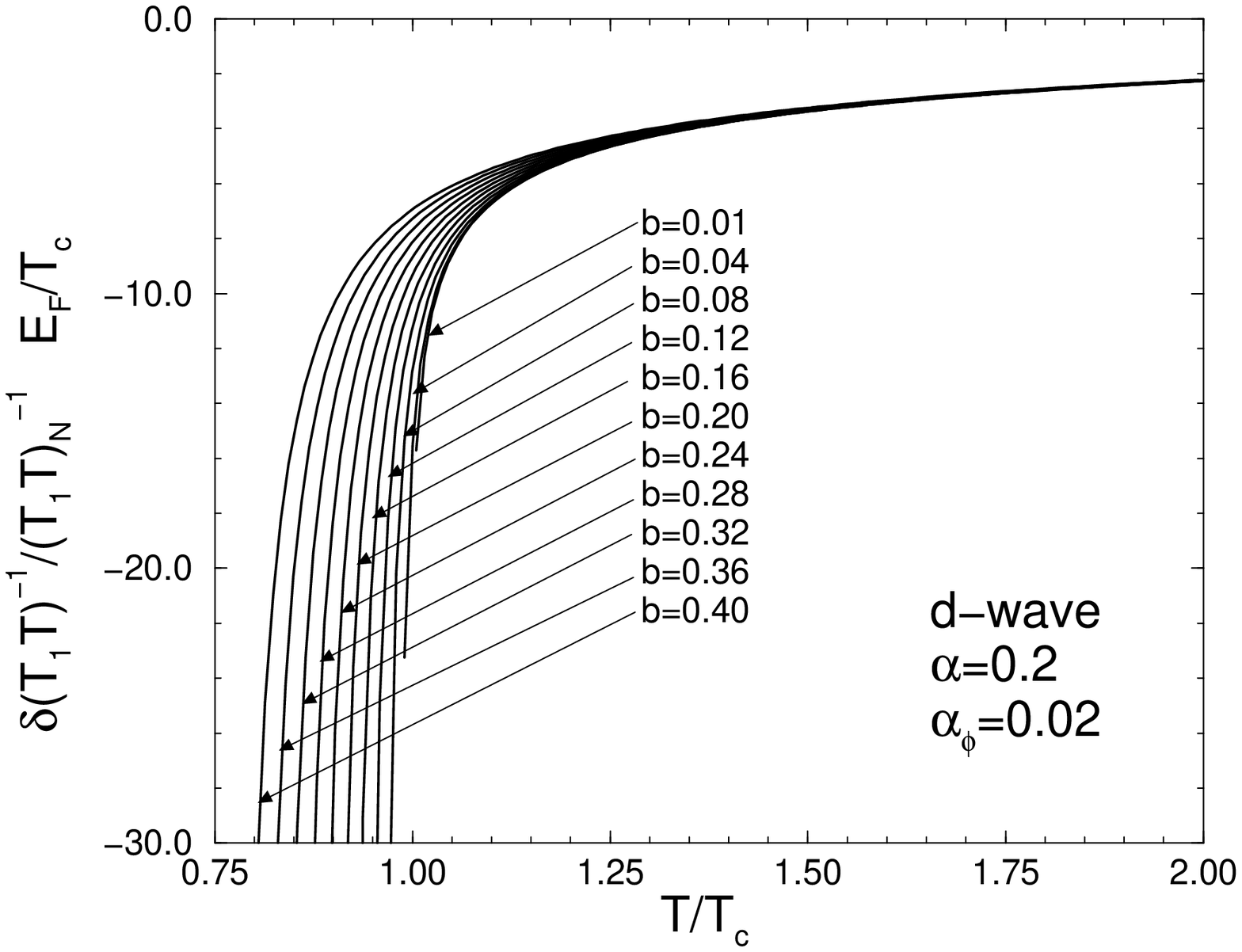}}
\begin{minipage}{0.95\hsize}
\caption{
\label{T1doverb2}
Temperature dependence of d-wave
fluctuation corrections to NSLR rate for different fields, given
as the sum of `anomalous Maki', `regular Maki' and `DOS' terms.}
\end{minipage}
\end{minipage}
\end{figure}

\begin{figure}
\begin{minipage}[t]{0.48\hsize}
\centerline{\epsfxsize0.93\hsize\epsffile{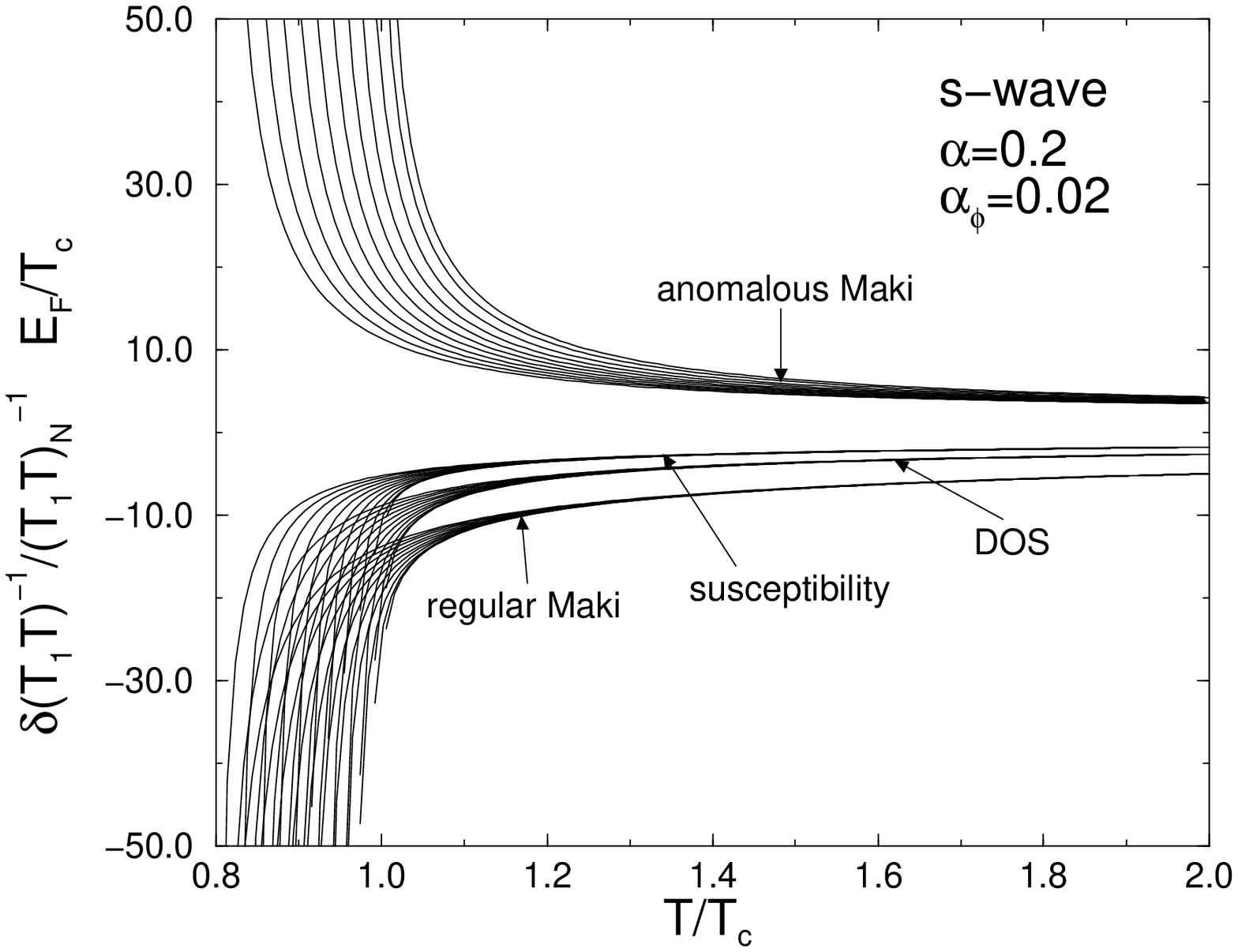}}
\begin{minipage}{0.95\hsize}
\caption[]{
\label{T1soverb1}
Temperature dependence of the Maki-\-Thomp\-son 
contributions and 
the DOS-contribution to the fluctuation corrections in the NSLR rate,
assuming s-wave pairing.
The curves are shown for different
fields, ranging
from 0.01, 0.04-0.4 (in steps of 0.04), from right to left.
For comparison, we also show the fluctuation corrections to the 
Pauli spin susceptibility. }
\end{minipage}
\end{minipage}
\begin{minipage}[t]{0.48\hsize}
\centerline{\epsfxsize0.93\hsize\epsffile{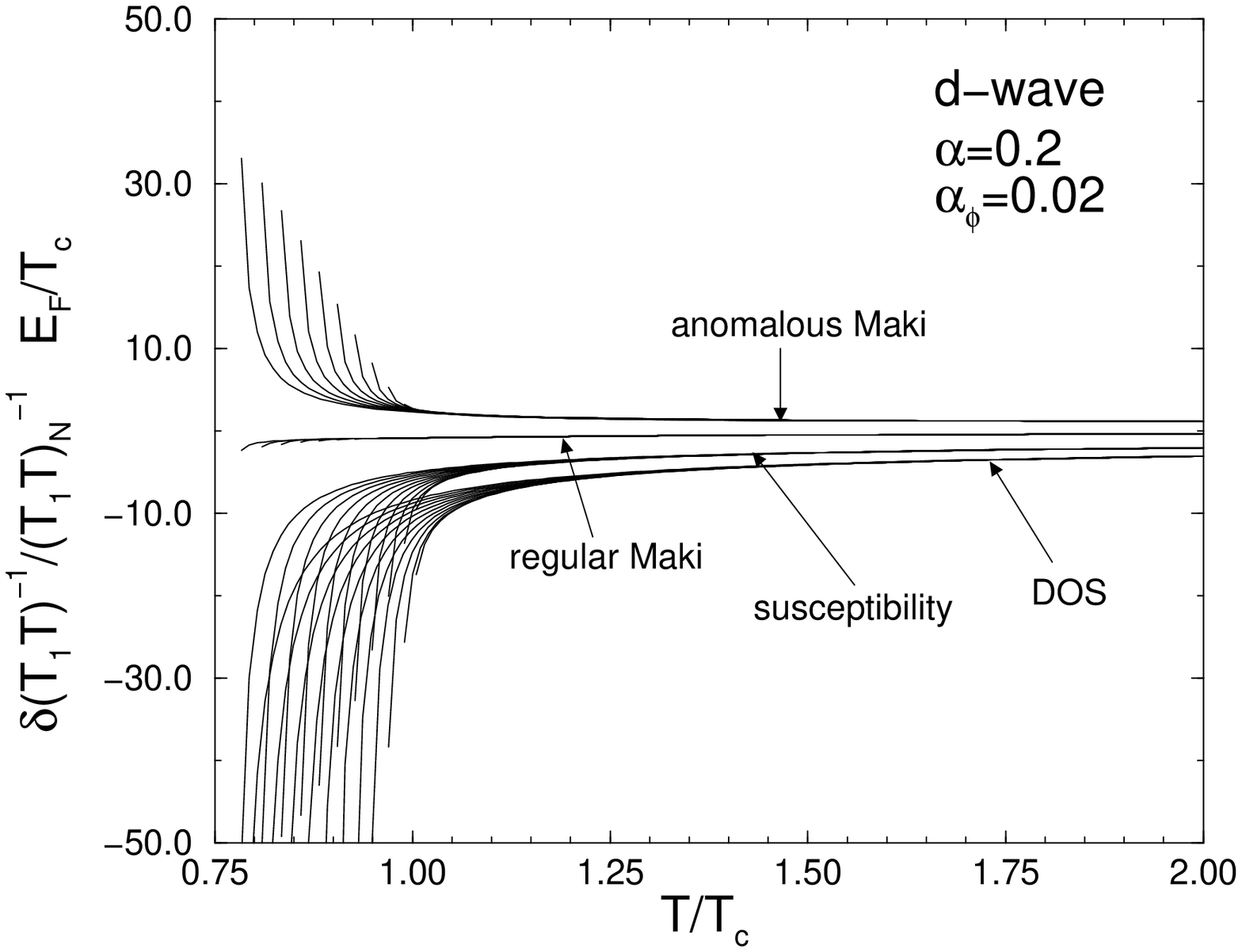}}
\begin{minipage}{0.95\hsize}
\caption[]{
\label{T1doverb1}
The same as in Fig. \ref{T1soverb1}, for d-wave pairing.
Note, that the regular Maki-Thompson term is negligible
compared to the other terms at all temperatures. The anomalous
Maki-Thompson term is
negligible at $b=0$ for $\alpha=0.2$ and $\alpha_\phi=0.02$, 
but contributes considerably at higher fields,
$b\gsim 0.2$.}
\end{minipage}
\end{minipage}
\end{figure}

The theory of leading order pairing fluctuations predicts
characteristic features in the temperature dependence of
the fluctuation corrections to the NSLR rate $1/T_1$. Typical  results
for s-wave and d-wave pairing are shown in Figs. \ref{T1s}-\ref{T1doverb1}.
For both symmetries 
there is a pronounced enhancement of the absolute value of the
fluctuation
corrections when the mean-field transition temperature, $T_c(b)$,
is approached. However, depending on the scattering parameters,
$\alpha$ and $\alpha_{\phi}$, the corrections may be positive
or  negative near $T_c(b)$. 

We first show in Figs. \ref{T1s}-\ref{T2d} the influence of impurities
on $1/T_1$ for small and intermediate values of the
magnetic field, $b=0.01$ and $b=0.4$.
For s-wave symmetry we show results for fixed elastic scattering,
$\alpha =0.2$, for a range of pair breaking parameters, $\alpha_\phi$.
In the low-field limit, shown in Fig. \ref{T1s},
a crossover from positive to negative divergence takes
place for $\alpha_\phi \approx 0.16$. The divergence is much weaker
for strong pair breaking compared to weak pair breaking in the
relatively clean case of $\alpha \lsim 0.2$ discussed in this paper.
Note that in the clean limit
the correction to $1/T_1$ diverges
like $\sqrt{T_c/(T-T_c)}$ in zero field,\cite{randeria94} 
compared
to a logarithmic divergence in the dirty case.\cite{kuboki89} In the dirty limit
the crossover
from a positive logarithmic divergence to a negative logarithmic divergence
takes place at $\alpha_\phi\approx 0.26$.\cite{alphacrit}

\end{multicols}
\begin{multicols}{2}
\noindent
The low-field results for
d-wave symmetry are shown in Fig. \ref{T1d}.
Because inelastic and elastic scattering act similarly in d-wave
superconductors we fixed $\alpha_\phi =0.001 $ and present results
for several values of the elastic scattering rate, $\alpha $. As can be seen
in Fig. \ref{T1d} there is a crossover from a positive to a negative
divergence for 
$\alpha \approx 0.04$, corresponding to a mean
free path of about 25 coherence lengths. 
For realistic values of scattering parameters in
high-$T_c$ superconductors, $\alpha +\alpha_\phi \approx 0.2$,
a negative divergence should be observed.

The effects of a strong field, $b=0.4$, are shown in Figs. \ref{T2s} and
\ref{T2d}.
For s-wave fluctuations the pair breaking effect
of the magnetic field dominates the effect of intrinsic
pair breaking, leading to large negative fluctuation
corrections to the NSLR rate as shown in Fig. \ref{T2s}. 
For d-wave pairing the effect of a magnetic field is much less
pronounced.
In the clean limit, even at high magnetic fields, the 
fluctuation contributions to the NSLR rate show a positive
divergence for d-wave pairing, in sharp contrast to s-wave pairing.
We discuss this result in more detail below.
However, for cuprate superconductors with d-wave pairing, and a reasonable
estimate for the scattering rate,
$\alpha \approx 0.2 $, we obtain a 
negative correction for all field strengths.

As can be seen by comparison of the NSLR rate for
$b=0.01$ and $b=0.4$, there
is a strong effect of the magnetic field on the
temperature dependence in s-wave superconductors.
The temperature dependences of the NSLR rate of superconductors
with s- and d-wave pairing are compared in Figs. \ref{T1soverb2}
and \ref{T1doverb2} for different magnetic field strengths
and parameters $\alpha=0.2$,
$\alpha_\phi=0.02$, which are typical estimates
for high-$T_c$ superconductors.\cite{tanner92}

In Fig. \ref{T1soverb2} we show, for s-wave pairing, that
there is a dramatic change in the behavior of the corrections
to the NSLR rate at field $b\approx 0.2 $ 
for $T$ near $T_c(b)$. Whereas for $b\lsim 0.2$ fluctuations
{\it enhance} the NSLR rate with decreasing temperature, 
for $b\gsim 0.2$ fluctuations {\it suppress} the NSLR rate with 
decreasing temperature. Note, that to observe this effect
one must compare the qualitative temperature behavior of the NSLR rate
for different fields rather than changing the magnetic field at constant
temperature.
For d-wave symmetry, shown in \mbox{Fig. \ref{T1doverb2}}, 
this effect is absent.

To clarify the origin of this behavior we have plotted
the Maki-Thompson terms, and the DOS term separately
in Figs. \ref{T1soverb1} and \ref{T1doverb1}.
As can be seen,
all contributions to the fluctuations are reduced in magnitude
at constant temperature with increasing magnetic field.
In contrast, 
all terms are enhanced in magnitude
with increasing magnetic field for constant $T-T_c(b)$,
as can be inferred from the larger slope of $\delta (T_1T)^{-1}$ near
$T_c(b)$ at lower fields.

Neither the DOS nor
the regular Maki-Thompson correction alone are large enough to 
dominate the anomalous Maki-Thompson contribution.
But together these two corrections overcompensate the
anomalous MT correction for fields above $b\gsim 0.2$
for s-wave pairing, which leads to the qualitative changes shown in
Fig. \ref{T1soverb2}.
In d-wave pairing the regular MT contribution
is negligible for all magnetic field strengths, as shown in Fig. \ref{T1doverb1}. 
This is true also for the regular MT contribution in the ultra-clean limit,
not shown here, and explains why
there is no change in sign of fluctuation corrections with
increasing magnetic field for d-wave pairing.
We also show for d-wave pairing
in Fig. \ref{T1doverb1}, that the anomalous Maki-Thompson term cannot be
neglected near $T_c$;
it diverges at the mean field transition temperature,
$T_c(b)$, except for zero magnetic field, $b$.

Finally, we suggest that the change
in sign of the fluctuation corrections to 
the NSLR rate for s-wave pairing
with increasing field should be observable in
the electron doped compounds like Nd$_{2-x}$Ce$_x$CuO$_{4-\delta}$, 
if they have s-wave pairing symmetry.
Observation of this effect would be a strong confirmation of s-wave
pairing in these compounds.

\end{multicols}
\begin{multicols}{2}

\subsection{Comparison with experiment}

In order to compare our results with experimental results obtained in
high-$T_c$ cuprate superconductors, we discuss first some specific
aspects of NMR in these compounds.
In addition to superconducting fluctuations 
antiferromagnetic spin fluctuations are believed to
 play an important role in the cuprates.\cite{millis}
A spin pseudogap may occur at the antiferromagnetic wave vectors
$\vec{q}=\vec{Q}_{AF}$, which manifests itself in the temperature
dependence of the 
NSLR rate of the Cu(2) nuclear spins.
The NSLR rate is proportional to the
slope at zero energy of the dynamical susceptibility at the
positions of the nuclei, i.e. $\lim_{\omega \to 0 }\chi''(\vec{R}_{\nu},
\omega)/\omega $, and is especially
sensitive to changes in the spectral weight of low-energy electronic 
excitations.
On the other hand, the Knight-shift
tensor, which probes the static spin susceptibility at $\vec{q}=0$,
is barely affected by the opening of the 
antiferromagnetic spin pseudogap at $\vec{Q}_{AF}$.
By contrast the opening of a pairing pseudogap at $\vec{q}=0$ affects
the quasiparticle density of states at the Fermi level, $N_F$,
and thus both the nuclear spin lattice relaxation rate ($\sim N_F^2$)
and the Knight shift ($\sim N_F$).

Recent experiments by Mitrovi{\'c} {\it et al.}\cite{mitrovic98}
and Bachman {\it et al.}\cite{bachman98}
reported the characteristic field scale on
which the pseudogap behavior
is suppressed, $H^\ast \approx $10 T in optimally doped YBCO.
Assuming that antiferromagnetic correlations lead to a 
suppression of spectral weight on the scale $J\sim 100 $ meV,
corresponding to $J/\mu_B \sim 1700 T$,
this comparatively low magnetic field scale has to be assigned to
another origin. 
Similarly, recent neutron scattering
experiments in fully oxygenated YBCO show that
the spin fluctuation spectrum near the antiferromagnetic wavevector
remains almost unaffected by a field of 11.5 Tesla.\cite{bourges97}
However, if spin fluctuations are responsible for the pairing
interaction between quasiparticles, it is possible that strong
coupling between quasiparticles and spin fluctuations may
lead to a pseudogap which has characteristics of both spin fluctuations
and pairing fluctuations.
At present a strong-coupling theory of superconducting fluctuations and 
antiferromagnetic spin
fluctuations has not been developed for pseudogap behavior in
high magnetic fields.

Our results are based on the theory of weak-coupling 2D pairing fluctuations.
Two-dimensional
fluctuation theory for YBCO is justified in the presence of
magnetic fields because of the large vortex liquid region below the
transition.
Phase coherence between planes may be neglected 
in the vortex
liquid state because of rapid thermal motion of the pancake vortices. Thus, it
is reasonable
to neglect the Josephson coupling
in the crossover region from the normal to vortex liquid
state as well.
This fact, and Landau level quantization in strong magnetic fields, 
implies that
fluctuations are predominantly two-dimensional.
It is possible that for fields smaller than 2 T a cross-over to
three-dimensional behavior might occur close to $T_c$.

\begin{figure}
\centerline{\epsfxsize0.93\hsize\epsffile{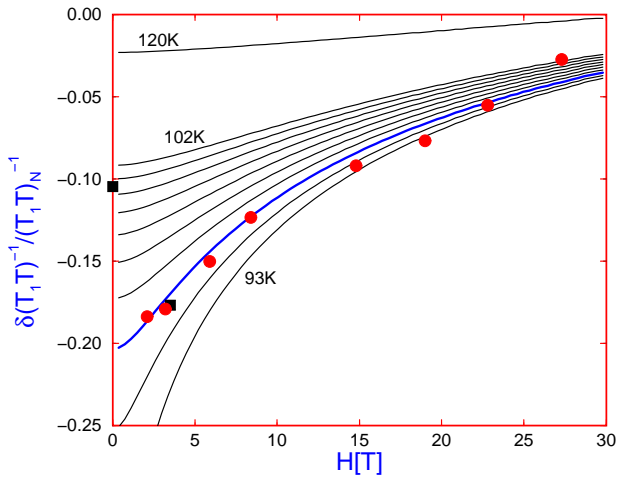}}
\begin{minipage}{0.95\hsize}
\caption[]{
\label{F4}
D-wave calculations for the superconducting pairing fluctuation contribution
$\delta (T_1T)^{-1}/(T_1T)^{-1}_N$  of $^{63}$Cu(2) 
spin-lattice relaxation rate in optimally doped YBCO
as a function of magnetic field
at temperatures ranging from 93 to 102 K in increments of 1 K, and for 120 K.
Circles\cite{mitrovic98} and squares\cite{song91}
are NMR and NQR (0 T) experiments.
The thick curve and the experimental data correspond to 95 K.
}
\end{minipage}
\end{figure}
We compare our calculations with experiments recently reported 
by Mitrovi{\'c} {\it et al.}\cite{mitrovic98} 
on optimally doped YBa$_2$Cu$_3$O$_{6.95}$ in a magnetic field $H||\hat c$.
Our calculations, which assume two-dimensional, d-wave pairing fluctuations
describe the experimental NMR data remarkably well.
The relative fluctuation correction $\delta(T_1T)^{-1}/(T_1T)^{-1}_N$
to $1/T_1T$ for d-wave pairing and $\alpha=0.2 $ and several temperatures
is shown in Fig. \ref{F4}.
We define the normal-state rate, $(T_1T)^{-1}_N$, to include
pairing fluctuation corrections 
that are constant in temperature and magnetic field.
Thus, to compare with experiments
we subtract these constant shifts from the calculated
fluctuation corrections as discussed in section \ref{magnmr}, and define
$\delta(T_1T)^{-1}\equiv(T_1T)^{-1}-(T_1T)^{-1}_N$. We chose the
value of the rate at 120K and 30T for this subtraction.
The experimental results from Mitrovi{\'c} {\it et al.}\cite{mitrovic98}
for the fluctuation correction are
also shown for the temperature $T= 95$K. 
In order to compare theory and experiment we subtracted from the
experimental data the asymptotic 
normal-state rate, which is well
described by $(T_1T)^{-1}_{N} \sim T_x/(T+T_x)$, to extract the
fluctuation correction $\delta (T_1T)^{-1}$.

The zero-field transition temperature of $T_c(0) = 92.5$ K
determines the absolute temperature scale for the theoretical calculations.
We solve numerically Eq. (\ref{hc2eq}) 
for the reduction of the mean-field transition temperature
as a result of Landau quantization.
Theoretically the mean-field transition temperature is
determined by diverging pairing fluctuations.
To fix the magnetic field scale we use the value for the mean field
transition temperature at 8.4 T
obtained from our fit to the fluctuation corrections to
the Pauli spin susceptibility,\cite{bachman98} discussed
in section \ref{Susc}.
There is one fitting parameter, $T_c/E_F$, which scales the magnitude of the
fluctuation contributions. As shown in Fig. \ref{F4}, the
agreement between the d-wave fluctuation theory 
and the experimental data from Ref. \onlinecite{mitrovic98} is excellent.

We also show in Fig. \ref{F4} data from Y.-Q. Song (black squares).\cite{song91}
The data point at $H=0$ is the NQR rate.
The NQR rate is {\it higher} than the low field NMR rate in the same
sample at 3.5 T.
A similar drop between the NQR rate and the low field NMR rate
was obtained by 
Carretta {\it et al.} on optimally doped YBCO.\cite{carretta96}
Based on the larger NQR rate compared with the NMR rate at
5.9 T, Carretta {\it et al.}
concluded that fluctuation corrections
to the NSLR rate are predominantly s-wave.\cite{carretta96} However, the field
evolution of the NSLR rate from 2 T to 27 T is in quantitative
agreement with the theory of 2D pairing fluctuations with d-wave
symmetry, and disagrees qualitatively and quantitatively with the
theory of s-wave fluctuations. The apparent discrepancy between
the NQR rate and the low-field NMR rate requires explanation.
We propose an explanation for the low-field evolution ($0 \le H \lsim 2$ T)
that reconciles Carretta {\it et al.}'s suggestion in terms of s-wave
pairing fluctuations with the field evolution and our
explanation in terms of d-wave pairing fluctuations.
We show below that sub-dominant s-wave fluctuations, induced by
the orthorhombic anisotropy of YBCO, can account
for the low-field evolution. At fields, $H\ge 2$ T the s-wave
fluctuations are suppressed and the dominant d-wave fluctuations
control the field evolution.

\subsection{Effect of orthorhombic distortion}
\label {orthorh}
If the crystal symmetry is not perfectly tetragonal, then the
s-wave and d-wave pairing channels correspond to the same
irreducible representation. Thus, the 
pairing basisfunction $\eta (\psi )$ is of the form
\begin{equation}
\eta (\psi ) = \beta_s \eta_s (\psi ) + \beta_d \eta_d (\psi )
\end{equation}
with $\beta_s^2+\beta_d^2=1$.
The results obtained for the fluctuation formulae 
for pure s- and d-wave pairing, Eqs. (\ref{rate})-(\ref{S_M}),
are the same with the replacements
\begin{eqnarray}
&&B_1(\epsilon,q,\phi)= 
\beta_d \frac{A_1(\epsilon,q,\phi)}{1-\tilde\alpha A_0(\epsilon,q)}+
\beta_s \frac{A_0(\epsilon,q,\phi)}{1-\tilde\alpha A_0(\epsilon,q)}, \\
&&B_2(\epsilon,q,\phi)= \beta_d^2 \frac{A_2(\epsilon,q,\phi) + \tilde\alpha
\big[ A_1^2-A_2A_0\big](\epsilon,q)}{1-\tilde \alpha A_0(\epsilon,q)} \nonumber \\
&&+\beta_s^2 \frac{A_0(\epsilon,q,\phi)}{1-\tilde\alpha A_0(\epsilon,q)}
+ 2 \beta_s \beta_d 
\frac{A_1(\epsilon,q,\phi)}{1-\tilde \alpha A_0(\epsilon,q)}.
\end{eqnarray}
Because $A_1 \sim \cos 2\phi $ the mixed terms in $B_2 $ and $B_1^2$ which
enter in the equations (\ref{S_D}) and (\ref{S_M}) are canceled to a large
extent by averaging over $\phi $. So near $T_c$ it is a good approximation
to add the s- and d-wave components of the
fluctuation corrections with weights $\beta_s^2 $ and
$\beta_d^2$, respectively.

\begin{figure}
\centerline{\epsfxsize0.93\hsize\epsffile{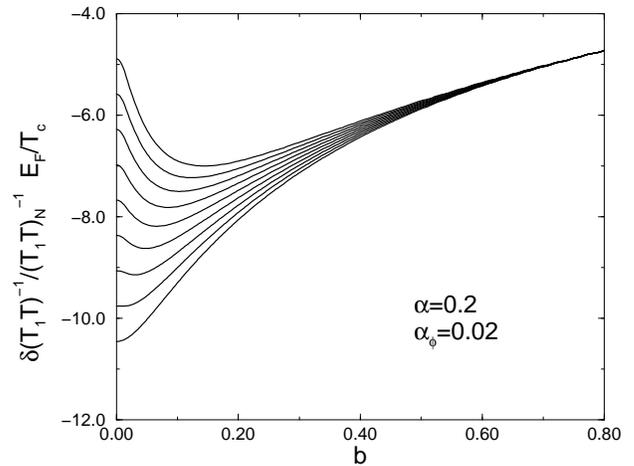}}
\begin{minipage}{0.95\hsize}
\caption[]{
\label{ortho}
Fluctuation corrections to nuclear spin relaxation rate taking into
account orthorhombic distortion. We assumed an induced asymmetry
in the order parameter described by
$\eta (\psi ) = \beta_d \eta_d (\psi ) + \beta_s \eta_s$, with
$\beta_s^2 $ varying from 0 to 0.2 in steps of 0.025 from
bottom to top, and $\beta_d^2=1-\beta_s^2$.  }
\end{minipage}
\end{figure}

In Fig. \ref{ortho} we show the result for 
$\delta (T_1T)^{-1} $ with $\beta_s^2$ ranging from 0 to 0.2.
Thus, the low field anomaly in the experimental data of Fig. \ref{F4}
can be accounted for by a small s-wave component induced by an orthorhombic
distortion, as can be seen comparing with Fig. \ref{ortho}.
We estimate $\beta_s^2\approx 0.15 $ ($\beta_d^2\approx 0.85$)
for optimally doped YBCO.
Note, that $b=0.8$ corresponds to $H=29$ T and that we account 
for both the position of the minimum in the NMR rate (at $\approx 2$ T)
and the difference between NQR and low field NMR rates
with one fitting parameter ($\beta_s$).

In BSCCO this 
effect should be absent if the dominant pairing channel has
B$_{1g}$-symmetry (d$_{x^2-y^2}$), because in this case the lattice distortion 
does not induce an s-wave, but rather a g-wave component with
A$_{2g}$ symmetry,
which has fluctuation corrections 
that respond to disorder and field similarly to the d-wave component.
However, an s-wave component would be induced if the order parameter of
BSCCO is predominantly B$_{2g}$-symmetry (d$_{xy}$).
\noindent

\section{Fluctuation corrections to the Pauli spin susceptibility}
\label{Susc}

The Pauli spin susceptibility is obtained from
the long wavelength limit of the particle-hole
susceptibility at $\omega_{e}=0$,
\begin{equation}
\chi_s= \mu_e^2 \sum_{k \alpha \beta }
\sum_{p \gamma \delta }
\left( \gvec{\sigma }_{\alpha \beta }\cdot \hat\vec{h} \right)
\left( \gvec{\sigma }_{\gamma \delta }\cdot \hat\vec{h} \right)
\chi^R_{p \gamma ,p \, \delta, k \alpha ,k \beta }(\omega_e=0)  ,
\end{equation}
where $\hat\vec{h}$ is a unit vector in direction of the applied field and
$\mu_e=\gamma_e\hbar/2 $.
The Pauli spin susceptibility
can be obtained from the spin part of the measured NMR Knight shift
by subtraction of the orbital and diamagnetic contributions.
Assuming an isotropic hyperfine matrix element,
$\mbox{$^{\nu}\vec{A}$}_{kk}$, and neglecting
anisotropic band structure and exchange interaction,
the spin shift, $K_{spin}$, is directly proportional to
the Pauli spin susceptibility, $\chi_s$.
The zeroth-order terms in $T_c/E_F$ for the particle-hole response
function (at $\omega_{e} =0$) define the Fermi-liquid result for the
Pauli spin susceptibility, $\chi_N$.

The spin susceptibility can be obtained directly from
the Matsubara Green's functions without analytic continuation
because it is an equilibrium quantity. 
Nevertheless, it is instructive to
write down the expression for $\delta \chi $ in terms of
retarded and advanced Green's functions 
defined on the real energy axis. 
\begin{figure}
\begin{minipage}{0.95\hsize}
\begin{center}
\begin{picture}(170,40)(0,0)
\SetScale{0.5}

\SetScaledOffset(10,30)
\SetScaledOffset(10,30)
\SetWidth{0.5} \Photon(-15,0)(0,0){2}{2.5}
\SetWidth{0.5} \Photon(-15,0)(0,0){-2}{2.5}
\SetWidth{2.5} \CArc(35,-5)(35,5,175)
\SetWidth{2.5} \CArc(35,5)(35,-175,-5)
\SetWidth{2.5} \CArc(105,-5)(35,5,175)
\SetWidth{2.5} \CArc(105,5)(35,-175,-5)
\SetWidth{0.5} \BCirc(70,0){15}
\SetPFont{Times-Roman}{15}
\PText(75,3)(-90)[]{K}
\SetPFont{Times-Roman}{10}
\SetPFont{Times-Roman}{10}
\SetWidth{0.5} \Line(43,-30)(43,30) \Line(27,-30)(27,30)
\SetWidth{0.5} \Line(113,-30)(113,30) \Line(97,-30)(97,30)
\SetPFont{Times-Roman}{15}
\PText(35,0)(0)[]{V}
\PText(105,0)(0)[]{V}
\SetPFont{Times-Roman}{10}
\SetWidth{0.5} \Photon(138,0)(153,0){2}{2.5}
\SetWidth{0.5} \Photon(138,0)(153,0){-2}{2.5}
\SetWidth{0.5} \GCirc(2,0){7}{0.7} \GCirc(138,0){7}{0.7}
\SetScaledOffset(165,30)
\Text(79,-3)[]{a)}

\SetScaledOffset(195,30)
\SetScaledOffset(210,30)
\SetWidth{0.5} \Photon(-15,0)(0,0){2}{2.5}
\SetWidth{0.5} \Photon(-15,0)(0,0){-2}{2.5}
\SetWidth{2.5} \Oval(60,0)(35,60)(0)
\SetWidth{2.5} \Oval(60,55)(15,20)(0)
\SetWidth{0.5} \BCirc(60,40){15}
\SetPFont{Symbol}{15}
\PText(58,43)(0)[]{K}
\SetPFont{Times-Roman}{10}
\SetPFont{Times-Roman}{10}
\SetWidth{0.5} \CArc(60,35)(23,-170,-10) \CArc(60,35)(37,-165,-15)
\SetWidth{0.5} \CArc(-10,-15)(30,-17,65) \CArc(-10,-15)(45,-20,57)
\SetWidth{0.5} \CArc(130,-15)(30,115,197) \CArc(130,-15)(45,123,200)
\SetPFont{Times-Roman}{15}
\PText(58,8)(0)[]{V}
\SetPFont{Times-Roman}{9}
\SetPFont{Times-Roman}{15}
\PText(25,-5)(0)[]{V}
\PText(95,-5)(0)[]{V}
\SetPFont{Times-Roman}{10}
\SetWidth{0.5} \Photon(120,0)(135,0){2}{2.5}
\SetWidth{0.5} \Photon(120,0)(135,0){-2}{2.5}
\SetWidth{0.5} \GCirc(0,0){7}{0.7} \GCirc(120,0){7}{0.7}
\SetScaledOffset(345,30)
\Text(170,-3)[]{b)}
\end{picture}\\
\begin{picture}(170,40)(0,0)
\SetScale{0.5}

\SetScaledOffset(10,30)
\SetScaledOffset(10,30)
\SetWidth{0.5} \Photon(-15,0)(0,0){2}{2.5}
\SetWidth{0.5} \Photon(-15,0)(0,0){-2}{2.5}
\SetWidth{2.5} \Oval(60,0)(35,60)(0)
\SetWidth{2.5} \Oval(60,-55)(15,20)(0)
\SetWidth{0.5} \BCirc(60,-40){15}
\SetPFont{Symbol}{15}
\PText(58,-37)(0)[]{K}
\SetPFont{Times-Roman}{10}
\SetPFont{Times-Roman}{10}
\SetWidth{0.5} \CArc(60,-35)(23,10,170) \CArc(60,-35)(37,15,165)
\SetWidth{0.5} \CArc(-10,15)(30,-65,17) \CArc(-10,15)(45,-57,20)
\SetWidth{0.5} \CArc(130,15)(30,-197,-115) \CArc(130,15)(45,-200,-123)
\SetPFont{Times-Roman}{15}
\PText(58,-3)(0)[]{V}
\SetPFont{Times-Roman}{9}
\SetPFont{Times-Roman}{15}
\PText(25,5)(0)[]{V}
\PText(95,5)(0)[]{V}
\SetPFont{Times-Roman}{10}
\SetWidth{0.5} \Photon(120,0)(135,0){2}{2.5}
\SetWidth{0.5} \Photon(120,0)(135,0){-2}{2.5}
\SetWidth{0.5} \GCirc(0,0){7}{0.7} \GCirc(120,0){7}{0.7}
\SetScaledOffset(525,30)
\Text(79,-3)[]{c)}

\SetScaledOffset(195,30)
\SetScaledOffset(210,30)
\SetWidth{0.5} \Photon(-15,0)(0,0){2}{2.5}
\SetWidth{0.5} \Photon(-15,0)(0,0){-2}{2.5}
\SetWidth{2.5} \Oval(60,0)(35,60)(0)
\SetWidth{2.5} \Oval(60,0)(30,20)(0)
\SetWidth{0.5} \BCirc(60,27){15}
\SetWidth{0.5} \BCirc(60,-27){15}
\SetPFont{Symbol}{15}
\PText(58,30)(0)[]{K1}
\PText(58,-23)(0)[]{K2}
\SetPFont{Times-Roman}{10}
\SetPFont{Times-Roman}{10}
\SetWidth{0.5} \CArc(-15,0)(30,-41,41) \CArc(-15,0)(45,-38,38)
\SetWidth{0.5} \CArc(135,0)(30,139,221) \CArc(135,0)(45,142,218)
\SetPFont{Times-Roman}{15}
\PText(22,3)(0)[]{V}
\PText(98,3)(0)[]{V}
\SetPFont{Times-Roman}{10}
\SetWidth{0.5} \Photon(120,0)(135,0){2}{2.5}
\SetWidth{0.5} \Photon(120,0)(135,0){-2}{2.5}
\SetWidth{0.5} \GCirc(0,0){7}{0.7} \GCirc(120,0){7}{0.7}
\SetScaledOffset(685,30)
\Text(170,-3)[]{d)}

\end{picture}
\end{center}
\vspace{0.5cm}
\begin{minipage}{0.95\hsize}
\caption[]{
\label{low}
First corrections in $T/E_F$ to the Pauli spin susceptibility.
$V$ denotes vertex corrections in the particle-hole channel;
$V=1$ in our model.
$K$, $K1$ and $K2$ denote the
(impurity renormalized) fluctuation modes in the pairing channel.
For the Pauli spin susceptibility
$K1$ and $K2$ either are a singlet Cooperon or
a triplet impurity Cooperon in a complementary way.}
\end{minipage}
\end{minipage}
\end{figure}
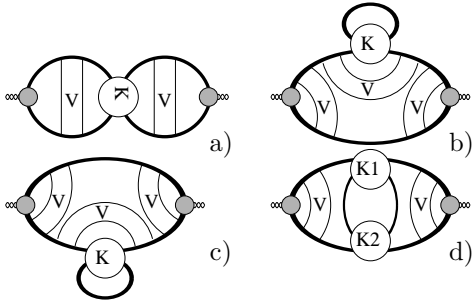

The pairing fluctuation corrections to leading order
in $T_c/E_F$ for the static ($\omega_{e}=0$) long wavelength ($q_{e}\to 0$)
spin susceptibility are obtained by the procedure discussed in
App. \ref{diag}, and are summarized by the 
DOS, Maki-Thompson
and Aslamazov-Larkin diagrams, shown in Fig. \ref{low}.\cite{ALdiagram} 
Note that in contrast to the large-$q_{e}$ response the contribution
d) in Fig. \ref{low} has the same order in $T_c/E_F$ as the DOS and
MT contributions.
However, it contains only one singlet pair fluctuation mode,
the other mode in the particle-particle channel is a triplet
impurity Cooperon mode.
Algebraic expressions for these diagrams are
given in Appendix \ref{Corr}. 
The sum of the leading order corrections in
Eq. (\ref{wsus}) of App. \ref{Corr}
leads to the following expression for the
relative fluctuation contribution to the Pauli spin susceptibility:
\begin{eqnarray}
\frac{\delta \chi }{\chi_N}=
\frac{T_c}{E_F} \int_0^{\infty} \frac{v_F^2 q dq}{2\pi T_c} \int_0^{2\pi}\frac{d\phi}{2\pi} S (q,\phi)
\end{eqnarray}
where $S (q,\phi)$ sums the contributions from all
diagrams shown in Fig. \ref{low} and is given by
\lrule
\begin{eqnarray}
S (q,\phi) &=&
\pi \int_0^{\infty} \frac{d\omega}{2\pi} \coth \frac{\omega}{2T}
\Re L(\omega,q,\phi) \int_0^{\infty}  \frac{d\epsilon}{2\pi}
\left[ \partial^2_\epsilon \tanh \frac{\epsilon+\omega/2}{2T} -
\partial^2_\epsilon \tanh \frac{\epsilon-\omega/2}{2T} \right] \Re 
B_2(\epsilon,q,\phi)  \nonumber \\
&+&\pi \int_0^{\infty} \frac{d\omega}{2\pi} \coth \frac{\omega}{2T}
\Im L(\omega,q,\phi) \int_0^{\infty}  \frac{d\epsilon}{2\pi}
\left[ \partial^2_\epsilon \tanh \frac{\epsilon+\omega/2}{2T} +
\partial^2_\epsilon \tanh \frac{\epsilon-\omega/2}{2T} \right] \Im 
B_2(\epsilon,q,\phi).
\end{eqnarray}
\rrule
Comparing with Eq. (\ref{S_D}) one realizes
that the fluctuation correction to the Pauli spin susceptibility
is given exactly by the first two lines of the density-of-states
contribution to the NSLR rate in Eq. (\ref{S_D}).
This result is nontrivial not only because the fluctuation corrections
to the NSLR rate and spin susceptibility are determined by different
diagrams, but particularly because the NSLR rate is a local response
defined by an integral over all wavelengths, while the spin susceptibility
is a global response obtained from the limit $q\to 0$.
The relative corrections to the spin susceptibility
as a function of the magnetic field
are shown in Figs. \ref{F1s} and \ref{F1d} (denoted by
`susceptibility'). 

\subsection{Results: Magnetic field and temperature dependence}

\begin{figure}
\begin{minipage}[t]{0.95\hsize}
\centerline{\epsfxsize0.93\hsize\epsffile{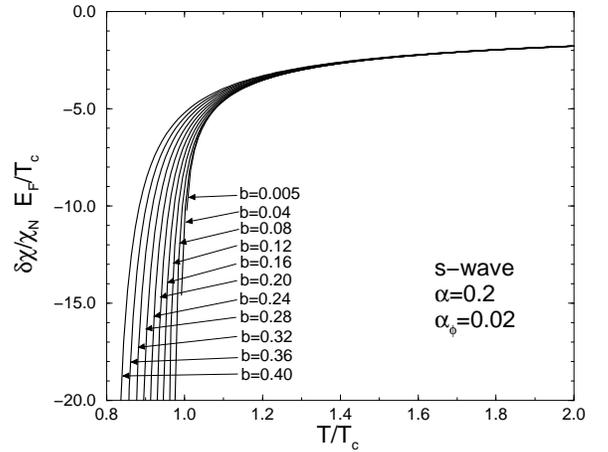}}
\begin{minipage}{0.95\hsize}
\caption[]{
\label{chisoverb1}
Temperature dependence of
fluctuation corrections to Pauli spin susceptibility for different fields.}
\end{minipage}
\end{minipage}
\end{figure}

\begin{figure}
\begin{minipage}[t]{0.95\hsize}
\centerline{\epsfxsize0.93\hsize\epsffile{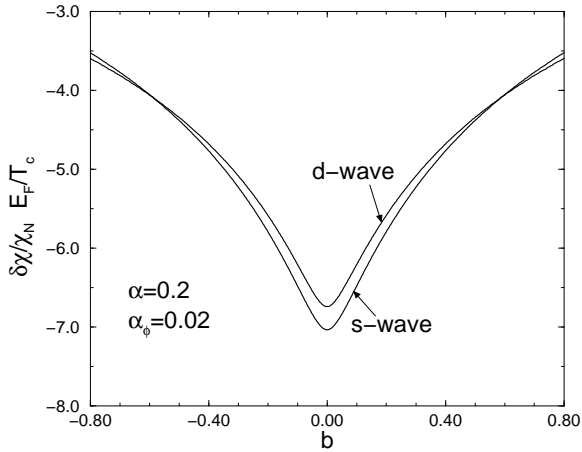}}
\begin{minipage}{0.95\hsize}
\caption{
\label{chisd}
Field dependence of fluctuation corrections to Pauli spin susceptibility
for $T/T_c$= 95K/92.5K $\approx 1.03$, and for s- and d-wave pairing.}
\end{minipage}
\end{minipage}
\end{figure}

Unlike the NSLR rate
the Pauli spin susceptibility is not very sensitive to either
impurity scattering or order parameter symmetry, as can be seen in Figs. \ref{F1s}
and \ref{F1d} for the magnetic field dependence of the susceptibility. 
Note that the small
constant offsets have to be subtracted off and are
included with the leading order terms as
discussed in section \ref{magnmr}.
The temperature dependence of the fluctuation corrections
to $\chi_s$ for s-wave pairing is shown in Fig. \ref{chisoverb1} for different
magnetic fields.
The shift in the divergence reflects the field dependence of $T_c(b)$.
In Fig. \ref{chisd} we show for comparison the magnetic field 
dependence of the fluctuation corrections
for s-wave and d-wave symmetry. 
As can be seen in this figure,
the fluctuation corrections to $\chi_s$ are insensitive
to the order parameter symmetry, at least for spin-singlet pairing
fluctuations.
Thus, the mixing of s- and d-wave
pairing fluctuations due to orthorhombic anisotropy,
which has a profound effect on the fluctuation
\noindent
corrections to the
NSLR rate at low field,
has almost no effect on the fluctuation corrections 
to the spin susceptibility.

\subsection{Comparison with experiment}

\begin{figure}
\centerline{\epsfxsize0.93\hsize\epsffile{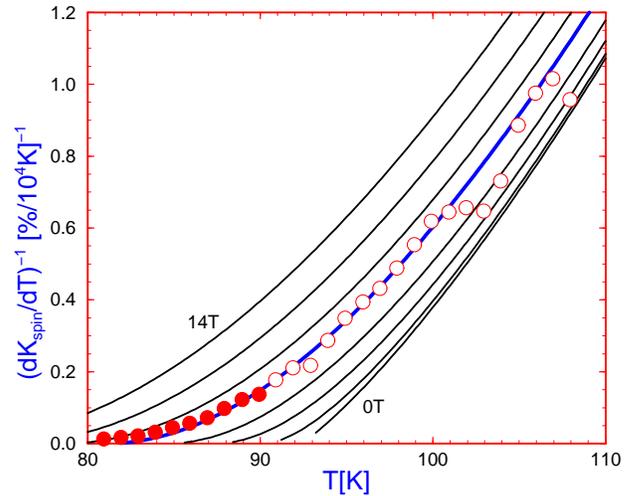}}
\begin{minipage}{0.95\hsize}
\caption[]{
\label{F5}
Calculations for 2D d-wave pairing fluctuation corrections to
the spin susceptibility for $B =$ 0 T - 14 T (in steps of 2 T).
Shown is the inverse of the derivative, $(d K_{spin}/d T)^{-1}$.
Circles show measurements of the
Knight shift of $^{17}$O(2,3) in optimally doped YBCO at 8.4 T.\cite{bachman98}
Open circles denote points used for the fit to $K_{spin}(T)$.
}
\end{minipage}
\end{figure}

Knight shift
measurements in high magnetic fields provide 
valuable information on the  fluctuation contributions to 
the Pauli spin susceptibility. 
The effect of static long-wavelength fluctuations on the Pauli
susceptibility in zero field have been calculated in
three dimensions\cite{fulde70} and two dimensions\cite{randeria94} 
for $\epsilon  = (T - T_{c})/T_{c} \ll 1$.
The fluctuation contribution
to the spin susceptibility was found to scale as
 $\delta \chi/\chi_N \sim \ln(\epsilon )$ in 2D, 
and $\delta \chi/\chi_N \sim \mbox{const}. + \sqrt{\epsilon }$ in
3D. For the 2D case one obtains
$(d \delta \chi/d T)^{-1} \sim T - T_{c} $, and
$(d \delta \chi/d T)^{-1}\sim \sqrt{T - T_{c}}$ for the 3D case.
Neither of these limiting cases is
consistent with the recent data of Bachman {\it et al.}\cite{bachman98}
on optimally doped YBCO shown in \mbox{Fig. \ref{F5}}.
These NMR measurements of the
Pauli spin susceptibility do not show singular behavior near
the transition.
This is typical for a
fluctuation-dominated crossover transition.\cite{bachman98} 
For this reason it is preferable to treat the 
mean-field transition temperature, $T_c(H)$, as a fitting parameter. 
The mean-field transition temperature was determined
by analyzing the high-precision  measurements of $^{17}$O(2,3) Knight shift
in optimally doped YBCO at high magnetic fields.\cite{bachman98}

The curvature shown in Fig. \ref{F5} is not reproduced by
2D static fluctuations in the low-field limit. Three dimensional fluctuations
to not account for the behavior because they produce curvature in
the opposite direction compared to the curves in Fig. \ref{F5}.
We can describe the behavior in Fig. \ref{F5} qualitatively and
quantitatively by taking into account 
dynamical fluctuations and orbital quantization.
The magnetic field is in a range where neither the low-field
approximation nor the lowest-Landau-level approximation is applicable.
We perform the sum 
\noindent
over the Landau levels and over the dynamical modes
numerically. Orbital quantization is the main source of the observed
curvature at higher fields. Dynamical fluctuations produce curvature
also for zero field, where orbital 
quantization is absent.

A quantitative comparison of our calculations with the experimental data
of Bachman {\it et al.}\cite{bachman98} is shown in Fig. \ref{F5}.
The fit was performed
in the region $T>90 $ K directly on the susceptibility data (open circles).
Then the inverse of the 
derivative of the experimental data and the theoretical curves
were calculated; they are extremely sensitive to variations at high
temperatures where the Pauli susceptibility deviates very little
from a constant. 
As can be seen in the Fig. \ref{F5}, the agreement is excellent even up to
temperatures of 102 K. 
The same fit accounts for the data in the non-fitted region (full circles)
down to 85 K.
The theoretical mean field temperature was determined
to be about 81 K at 8.4 T.
As we discussed in Fig. \ref{chisd}, mixing of an s-wave contribution
due to orthorhombic anisotropy in YBCO has
little influence on the fluctuation corrections to
the Pauli spin susceptibility.

\section{Conclusions}
We have calculated the pairing fluctuation corrections to the
nuclear spin-lattice relaxation rate and to the Pauli spin
susceptibility in 2D s-wave and d-wave high-$T_c$ superconductors in
strong magnetic fields. Our calculations include 
dynamical and short wavelength fluctuations.
We account qualitatively and quantitatively for
recent experiments performed on optimally doped YBCO solely in
terms of d-wave pairing fluctuations, assuming reasonable scattering
parameters.
We find no 
necessity to invoke the existence of a spin-density fluctuation pseudogap.
We have shown that incorporating orthorhombic anisotropy
and the allowed mixing of s-wave and d-wave pairing fluctuation channels
leads to a low-field crossover from predominantly s-wave fluctuations
to predominantly d-wave fluctuations which provides a natural
explanation for the observed evolution from the NQR rate to the
low-field (below 2 Tesla) $^{63}$Cu NSLR rate 
on optimally doped YBCO.
We suggest that a change in sign of the fluctuation corrections to 
the NSLR rate near $T_c(H)$ with increasing field should be observable in
the electron doped compounds like Nd$_{2-x}$Ce$_x$CuO$_{4-\delta}$, 
if they have s-wave pairing symmetry.
Observation of this effect would be a strong confirmation of s-wave
pairing in these compounds.

\acknowledgements
We are particularly thankful to V. F. Mitrovi{\'c}, H. N. Bachman,
W. P. Halperin, and Y.-Q. Song for providing us with 
experimental data prior publication.
We gratefully acknowledge useful discussions with
M. Fogelstr\"om, J. Heym, and S.-K. Yip.
This work is supported by the National Science Foundation (DMR 91-20000) 
through the Science and Technology Center for Superconductivity.
DR and JAS also acknowledge support from the Max-Planck-Gesellschaft
and the Alexander von Humboldt-Stiftung.
ME also acknowledges support from the Deutsche Forschungsgemeinschaft.
\lrule
\begin{appendix}
\rrule
\section{Irreducible pair susceptibilities for d-wave symmetry }
\label{Afunctions}

In this section we summarize expressions for the
$\xi$-integrated Fermi-surface averages of the 
product Green's functions at real energies for the case of
d-wave pairing in 2D, i.e. $\eta (\psi )= \sqrt{2}\cos 2\psi$. 
The integrals are related to Eq. (\ref{intgf}) by analytic continuation.
For d-wave pairing they are,
\begin{eqnarray}
\label{A0}
&&A_0(\epsilon,q)=\frac{2\pi N_F}{\sqrt{(v_Fq)^2-(2\tilde \epsilon )^2}}, \\
\label{A1}
&&A_1(\epsilon,q,\phi)=A_0(\epsilon,q) \frac{-2i\tilde \epsilon - \sqrt{(v_Fq)^2-(2\tilde \epsilon )^2}
}{-2i\tilde \epsilon + \sqrt{(v_Fq)^2-(2\tilde \epsilon )^2}}
\sqrt{2} \cos 2\phi,  \nonumber \\ \\
&&A_2(\epsilon,q,\phi)= \nonumber \\
&&A_0(\epsilon,q) \left[ 1 + \left( \frac{-2i\tilde \epsilon - \sqrt{(v_Fq)^2-(2\tilde \epsilon )^2}
}{-2i\tilde \epsilon + \sqrt{(v_Fq)^2-(2\tilde \epsilon )^2}}\right)^2
\cos 4\phi \right] ,\\
&&\big[ A_1^2-A_2A_0\big](\epsilon,q)= \nonumber \\
&&-A_0(\epsilon,q)^2 
\left[ 1 - \left( \frac{-2i\tilde \epsilon - \sqrt{(v_Fq)^2-(2\tilde \epsilon )^2}
}{-2i\tilde \epsilon + \sqrt{(v_Fq)^2-(2\tilde \epsilon )^2}}\right)^2 \right],
\end{eqnarray}
with $\tilde \epsilon = \epsilon + i\pi T_c (\alpha +\alpha_\phi) $.

For isotropic s-wave pairing fluctuations we have trivially
$A_0(\epsilon,q)=A_1(\epsilon,q)=A_2(\epsilon,q)$ given in
(\ref{A0}).

\section{Coherence lengths and $T_c$-reduction}
\label{coherence}
In the long-wavelength, low-frequency limit
the pair fluctuation propagator
for s- and d-wave symmetry becomes\cite{varlamov98}
\begin{eqnarray}
L_s(q,\omega)&=&N_F^{-1}\frac{1}{\epsilon_s +\xi_s^2 q^2 -i\omega \tau_s}\\
L_d(q,\omega)&=&N_F^{-1}\frac{1}{\epsilon_d +\xi_d^2 q^2 -i\omega \tau_d}.
\end{eqnarray}
This result
is obtained by expanding Eq. (\ref{Lreal}) for small $q$ and $\omega $ and
carrying out the $\epsilon $-integral.
We define $\alpha_0=\alpha+\alpha_\phi$,
then,
\begin{eqnarray}
\frac{\xi_s^2}{\xi_0^2}&=& \frac{\Psi\big(\frac{1}{2}+\frac{\alpha_\phi T_c}{2T}\big)-\Psi\big(\frac{1}{2}+\frac{\alpha_0 T_c}{2T}\big)+
\frac{\alpha T_c}{2T}\Psi'\big(\frac{1}{2}+\frac{\alpha_\phi T_c}{2T}\big)}{2\alpha^2}, \nonumber \\ \\
\frac{\xi_d^2}{\xi_0^2 }&=& \frac{T_c^2}{T^2}\; \frac{\left|\Psi''\big(\frac{1}{2}+\frac{\alpha_0 T_c}{2T}\big)\right|}{16},\\
\tau_s&=& \frac{\Psi'\big(\frac{1}{2}+\frac{\alpha_\phi T_c}{2T}\big)}{4\pi T},\\
\tau_d&=& \frac{\Psi'\big(\frac{1}{2}+\frac{\alpha_0 T_c}{2T}\big)}{4\pi T},\\
\epsilon_s&=&\ln \frac{T}{T_c} 
-\Psi\left(\frac{1}{2}+\frac{\alpha_\phi}{2}\right)+
\Psi \left(\frac{1}{2}+\frac{\alpha_\phi T_c}{ 2T}\right),\\
\epsilon_d&=&\ln \frac{T}{T_c} 
-\Psi\left(\frac{1}{2}+\frac{\alpha_0}{2}\right)+\Psi \left(\frac{1}{2}+\frac{\alpha_0 T_c}{2T}\right).
\end{eqnarray}
Note that s- and d-wave results differ by more than the
replacement of $\alpha_\phi $ by $\alpha_\phi + \alpha $. The
relative influence of $\alpha $ and $\alpha_\phi$ on the
reduction of the coherence length is different for d-wave and
s-wave symmetry.

The reduction of $T_c$ by impurity scattering is given by the 
Abrikosov-Gorkov formulas\cite{Abgo}
\begin{eqnarray}
\ln \frac{T_{c}}{T_{c0}} -\Psi\left(\frac{1}{2}\right)+
\Psi \left(\frac{1}{2}+\frac{\alpha_\phi }{ 2}\right)
&=&0 \quad \mbox{(s-wave)}\\
\ln \frac{T_{c}}{T_{c0}} -\Psi\left(\frac{1}{2}\right)+
\Psi \left(\frac{1}{2}+\frac{\alpha_0 }{ 2}\right)
&=&0 \quad \mbox{(d-wave)}.
\end{eqnarray}

\section{Classification of diagrams}
\label{diag}
The essential feature for our classification scheme of diagrams
is a separation of energy scales. 
The {\em low energy scale} set by the temperature ($k_BT$),
the quasiparticle excitation energy ($\epsilon$), 
the pair excitation energy ($\omega $), the scattering rates
($\hbar/\tau$, $\hbar/\tau_\phi$), etc.
should be well separated from the characteristic {\em high energy scales}
of the metal, e.g. the Fermi energy ($E_F$).
These energies define a formal
expansion parameter given by the ratio of a typical
low energy scale and a typical high energy scale, for instance,
$k_BT_c/E_F$.
Alternatively one can
write the formal expansion parameter in terms of the ratio of
a typical atomic length scale ($k_F^{-1}$, $\hbar v_F/E_F$, etc.) and a
typical long wavelength scale
($\xi_0=\hbar v_F/2\pi k_BT_c$, $\ell=v_F\tau, \ell_\phi=v_F\tau_{\phi}$, etc).
We perform a systematic expansion in terms of these parameters, and
derive all leading 
fluctuation corrections in the framework of the Green's functions technique.
All diagrams presented here are understood as containing
renormalized elements. Thus, low-energy fermion Green's functions are
{\it quasiparticle}-Green's functions, vertices are renormalized by
high-energy quantities.
More detailed descriptions of this renormalization procedure are given in
Refs. \onlinecite{SeRai83,Rai86,Rai95}.
We assign for simplicity the  order of magnitude $small$ to the set of
expansion parameters (e.g. $small= T_c/E_F$).
To estimate the order of magnitude of the diagrams 
we replace the Green's functions for the
quasiparticles by piecewise constant functions, which are
equal to $1/small$ if both the momenta are located
in a narrow shell of thickness $small$
around the Fermi surface and the  energies are small, $\epsilon < small$.
The corresponding part of phase space is called
{\it low-energy region}.
Outside of this phase space area, in the {\it high-energy region}, we
assign to the phase space area a measure of one, and
the high-energy Green's functions are set equal to one.
Analogously, the low-energy range of a pair fluctuation mode consists of 
small pair excitation energies, $\hbar \omega < small$, and
small pair momenta, $|\vec{q}|<small$.
Performing the trivial integrations over the step-like Green's functions in the
asymptotic limit $small \to 0$ gives the order of the diagram.
This is done in the following steps:
\begin{enumerate}
\item Estimate the integrand from
the number of quasiparticle lines in the diagram, $n_Q$, which gives a factor
$small^{-(n_Q)}$.
\item { Labeling of the diagram respecting energy and momentum conservation.}
\item { Estimate the phase space factors:}\\
a)  Restricting all energies to their low-energy region
gives a factor $small^{(n_E)}$, where, $n_E$ is
the number of independent internal energies.\\
b)  Restricting the pair momentum to its low-energy region gives
a factor $small^D$ for every quasiparticle pair in the fluctuation channel, 
which is not otherwise restricted to the low energy region. 
The physical dimension, $D$, enters explicitly.\\
c) Restricting all remaining fermion momenta to their low-energy region
is the only nontrivial part of the estimate.
The number of restrictions, $n_K$, gives a factor $small^{(n_K)}$.
Note, that the sum of two low-energy momenta is not necessarily in the
low-energy region again. One needs additional geometrical
restrictions to the angles between the momenta.
\end{enumerate}
The leading order corrections in 2D to the NSRL rate are
determined by the diagrams in Fig.  \ref{high} 
for short-wavelength external perturbations, $\hbar q_e \sim p_F$, while
the leading order corrections in 2 D to the spin susceptibility
are determined by the diagrams in Fig. \ref{low} for long-wavelength
external perturbations, $\hbar q_e \ll p_F$.

In three dimensions
these corrections are another order higher in $small$, showing the
insignificance of fluctuations in conventional 3D superconductors.
In one dimension
they are of leading order, signaling the breakdown of the quasiparticle
picture.

\section{Corrections to the particle-hole susceptibility}
\label{Corr}

We use a short-hand notation, for the
combined (bosonic) Matsubara-energy and momentum
of the the pairing fluctuation mode: $Q \equiv (\omega_l, \vec q)$.
Similarly, $P \equiv (\epsilon_n, \vec p )$,
$P' \equiv (\epsilon_{n'}, \vec p\,')$, $Q-P \equiv
(\omega_l-\epsilon_n, \vec q -\vec p )$,
$\sum_{P} \equiv T\sum_n \sum_{\vec{p}}$ etc.
We use the usual Feynman rules for evaluating diagrams.\cite{Abgo}
Although we consider spin-singlet, s- or d-wave pairing,
both spin-singlet and spin-triplet fluctuation channels contribute
because of triplet impurity Cooperons.
We neglect the Zeeman coupling of the quasiparticle propagators
to the magnetic field. 
This allows us to decompose
the vertices and fluctuation propagator in the particle-particle
channel into spin-singlet and spin-triplet components,
\begin{eqnarray}
\label{sym1}
\displaystyle
&&\Gamma_{\alpha \beta \gamma \delta }(P,P',Q)= 
\nonumber \\ &&
\Gamma^s(P,P',Q)
\, \sigma^y_{\alpha \beta }\sigma^y_{\gamma \delta }
+\Gamma^t(P,P',Q)\,
(\sigma^y\! \gvec{\sigma })_{\alpha \beta }
(\gvec{\sigma }\sigma^y)_{ \gamma \delta } \, \, ,\\
\label{Ksym1}
\displaystyle
&&K_{\alpha \beta \gamma \delta }(P,P',Q)= 
\nonumber \\ &&
K^s(P,P',Q)
\, \sigma^y_{\alpha \beta }\sigma^y_{\gamma \delta }
+K^t(P,P',Q)\,
(\sigma^y\! \gvec{\sigma })_{\alpha \beta }
(\gvec{\sigma }\sigma^y)_{ \gamma \delta } \, \, .
\end{eqnarray}
The Bethe-Salpeter equation for the fluctuation propagator,
\lrule
\begin{eqnarray}
\label{sym2}
&&K_{\alpha \beta \gamma \delta }(P,P',Q)=
\Gamma_{\alpha \beta \gamma \delta }(P,P',Q)+
\frac{T}{2} \sum_{\epsilon ,\eta } \sum_{P''}
\Gamma_{\alpha \beta \eta \epsilon }(P,P'',Q)
\cdot
G(P'') G(Q-P'') \cdot
K_{\epsilon \eta \gamma \delta }(P'',P',Q) \, \, .
\end{eqnarray}
\rrule
\noindent
separates into singlet and triplet channels,
\begin{eqnarray}
\label{sym3}
\displaystyle
&&K^{s,t}(P,P',Q)=
\Gamma^{s,t}(P,P',Q)+
T\cdot \sum_{P''} \,\Gamma^{s,t}(P,P'',Q)
\nonumber \\ &&\times
G(P'') G(Q-P'') \cdot
K^{s,t}(P'',P',Q) \, \, .
\end{eqnarray}

Corrections to the NSLR rate are described by the diagrams in Fig. \ref{high}.
The first diagram was investigated by Maki and Thompson,\cite{CarMakThomp67}
and the last two diagrams represent contributions to the NSLR
rate from fluctuation contributions to the quasiparticle density of states.
Particle-hole vertex corrections, labeled $V$ in Fig. \ref{high},
can be neglected to leading order in $T_c/E_F$ above $T_c$ because
they are all proportional to
$\int d\xi_{\vec k} \, G(\epsilon_n ,\xi_{\vec k} )^2 \approx 0$.
The expressions corresponding to the diagrams in Fig. \ref{high},
with external Matsubara energy $\omega_m$, are then (we use
$W\equiv (\omega_m,\vec{q}')$)
\lrule
\begin{eqnarray}
\chi_{MT}(\omega_m)&=&
\sum_{\alpha \beta \gamma \delta } 
\sum_{PQ\vec q' } \,
 (\gvec{\sigma }_{\delta \alpha} \vec{A}_{-P, W-P } )\,
(\gvec{\sigma }_{\gamma \beta } \vec{A}_{P,P-W} )  \,
G(P -W)\, G(P)
\, G(Q -P ) \, G(Q -P + W ) \,
 K_{\alpha \beta \gamma \delta }(P,P-W ,Q) \nonumber \\
&=&
-2 
\sum_{P Q\vec q' } \,
|\vec{A}_{P,P-W}|^2  \,
G(P -W)\, G(P) \, G(Q -P ) \, G(Q -P + W ) \,
\left( K^s(P,P-W,Q)-K^t(P,P-W,Q)\right)  \, , \nonumber \\
\\
\chi_{DOS}(\omega_m)&=&
\sum_{\alpha \beta \gamma }
\sum_{P Q \vec q'} \,
(\gvec{\sigma }_{\alpha \gamma } \vec{A}_{P-W,P }) \,
(\gvec{\sigma }_{\gamma \alpha } \vec{A}_{P,P-W })  \,
\big( G(P -W )+ G(P+W) \big) \, G(P )^2\, G(Q -P ) \,
K_{\alpha \beta \beta \alpha }(P,P,Q ) \nonumber \\
&=& 4
\sum_{P Q \vec q'} \,
|\vec{A}_{P-W,P }|^2 \,
\big( G(P -W )+ G(P+W) \big) \, G(P )^2\, G(Q -P ) \,
\left( K^s(P,P-W,Q)+3K^t(P,P-W,Q)\right)  \, .\nonumber \\
\end{eqnarray}
\rrule
The term $\chi_{MT}$ corresponds to the Maki-Thompson diagram, a) in
Fig. \ref{high}, and the second term, $\chi_{DOS} $, to the two DOS diagrams,
b) and c) in Fig. \ref{high}.
We use the relations
$\vec{A}_{-P,-P' } = (\vec{A}_{P,P'})^\ast $ and
$\vec{A}_{P',P } = (\vec{A}_{P,P'})^\ast $ to simplify the results.

The fluctuation corrections to the Pauli
spin susceptibility are obtained from the diagrams in Fig. \ref{low},
\lrule
\begin{eqnarray}
\chi_{MT}(0)&=& 
\sum_{\alpha \beta \gamma \delta } 
\sum_{P Q }\,
(\vec{h}\cdot \gvec{\sigma }_{\delta \alpha }) \,
(\vec{h} \cdot \gvec{\sigma }_{\gamma \beta })  \,
G(P )^2 \, G(Q -P )^2
K_{\alpha \beta \gamma \delta }(P,P ,Q ) \nonumber \\
&=& -2\sum_{P Q } \,
G(P )^2 \, G(Q -P )^2
\left( K^s(P,P,Q)-K^t(P,P,Q)\right) \, , \\
\chi_{DOS}(0)&=& 
2 \sum_{\alpha \beta \gamma }
\sum_{P Q }\,
(\vec{h}\cdot \gvec{\sigma }_{\alpha \gamma }) \,
(\vec{h}\cdot \gvec{\sigma }_{\gamma \alpha })  \,
G(P )^3\, G(Q -P )
K_{\alpha \beta \beta \alpha }(P,P,Q ) \nonumber \\
&=& 4 \sum_{P Q }\,
G(P )^3\, G(Q -P )
\left( K^s(P,P,Q)+3K^t(P,P,Q)\right) \,  , \\
\chi_{AL}(0)&=& 
\sum_{\alpha \beta \gamma \delta \eta \zeta }
\sum_{P P' Q }
( \vec{h} \cdot \gvec{\sigma }_{\zeta \alpha } ) \,
( \vec{h} \cdot \gvec{\sigma }_{\eta \beta } ) \, 
G(P )^2 \, G(P' )^2\, G(Q -P ) \, G(Q -P'  )
K_{\alpha \beta \gamma \delta }(P ,P',Q )\,
K_{\delta \gamma \eta \zeta }(P' ,P ,Q ) \nonumber \\
&=& 4 \sum_{P P' Q }\,
G(P )^2 \, G(P' )^2\, G(Q -P ) \, G(Q -P'  )
\left( K^s(P,P',Q)K^t(P',P,Q)+ K^t(P,P',Q)K^s(P',P,Q) \right) \, .
\end{eqnarray}
\rrule
The first term, $\chi_{MT}(0)$, corresponds to diagram a) in Fig. \ref{low} 
(Maki-Thompson), the second term, $\chi_{DOS}(0)$, 
to diagrams b) and c) (DOS) and the last
term, $\chi_{AL}(0)$, to the Aslamazov-Larkin diagram, d) in Fig. \ref{low}.
Particle-hole vertex corrections, labeled $V$ in Fig. \ref{low},
can be neglected for similar reasons as in the case of the NSLR rate.

To evaluate momentum integrals 
we split the $\vec{p} $-sum into a $\xi_{\vec{p}} $-integral
and a Fermi surface average (we
 use the notation $\langle \ldots \rangle_{\vec p} := \int \, d p_F \, n(p_F) \ldots $
where $n(p_F) $ is the angle resolved (normalized) density of states at the Fermi
 surface). Thus,
\begin{eqnarray}
\sum_{\vec{p} } &\approx & N_F \cdot \int_{-\infty }^{\infty } d\xi_{\vec{p} } \cdot \langle \ldots
\rangle_{\vec p} \, \, .
\end{eqnarray}
The lower limit of the integrals is extended from $-\mu $ to $-\infty $. 
This approximation induces corrections of order $T_c/E_F$ which vary
on a temperature (and field) scale large compared to $T_c$ and can
be incorporated in the asymptotic normal state behavior as discussed
in section \ref{magnmr}.
We use the abbreviations
\begin{eqnarray}
\label{G1}
\underline G_1(P)&=&N_F \int d\xi_{\vec{p} } \, G(P) \\
\label{G2}
\underline G_2(P,Q)&=& N_F \int d\xi_{\vec{p} } \, G(P)G(Q-P) \\
\label{G3}
\underline G_3(P,Q)&=& N_F \int d\xi _{\vec{p} } \, G(P)^2G(Q-P) \\
\label{G3a}
\underline G_{3'}(P,Q)&=& N_F \int d\xi _{\vec{p} } \, G(P)G(Q-P)^2 \\
\label{G4}
\underline G_4(P,Q)&=& N_F \int d\xi_{\vec{p} } \, G(P)^3G(Q-P) \\
\label{G4a}
\underline G_{4'}(P,Q)&=& N_F \int d\xi_{\vec{p} } \, G(P)^2G(Q-P)^2.
\end{eqnarray}
These expressions can be evaluated by complex integration.
After $\xi$-integration 
(as on the left hand sides of the above equations) the momenta 
are confined to the Fermi surface.
The formal identity
\begin{equation}
\Big( G[\Sigma(\epsilon_n)](1) \Big)^{m+1} = \frac{1}{m!} \left( \frac{\delta}{\delta \Sigma(\epsilon_n)}\right)^m
G[\Sigma(\epsilon_n)](P)
\end{equation}
where $\Sigma(\epsilon_n)$ is the self energy for the Green's function,
implies
\begin{eqnarray}
&&\underline G_{2+m}[\Sigma(\epsilon_n),\Sigma(\omega_l-\epsilon_n)](P,Q)= \nonumber \\
&&\frac{1}{m!} 
\left( \frac{\delta}{\delta \Sigma(\epsilon_n)}\right)^m
\underline G_2[\Sigma(\epsilon_n),\Sigma(\omega_l-\epsilon_n)](P,Q).
\end{eqnarray}
Because the functional dependence of $\underline G_2$ on the self energies 
contains only the combination $\Sigma(\epsilon_n)-\Sigma(\omega_l-\epsilon_n)$, we obtain the relations
$\underline G_{4'}=-2\underline G_4$ and $\underline G_{3'}=-\underline G_3$.

In the weak-coupling theory for pair fluctuations we have
$K^s(P,P',Q)\equiv K(P,P',Q)=\tilde \eta (P,Q) L(Q) \tilde \eta (P',Q)$,
and we can 
replace $K^t(P,P',Q)$ and $K^t(P',P,Q)$ in the Aslamazov-Larkin diagram by
$C(\epsilon_n,Q) [\delta (\epsilon_{n'}-\epsilon_n)-\delta( \omega_l-\epsilon_{n'}-\epsilon_n)]/2$.
The quantities $\tilde \eta $, $L$ and $C$ are defined in section \ref{pairpropagator}.
We neglect diagrams containing only impurity interactions (and no 
pairing interaction), which describe pure weak localization effects.
Furthermore, we assume that the hyperfine matrix elements are isotropic
on the Fermi surface.
Thus, we obtain for the NSLR rate,
\lrule
\begin{eqnarray}
\chi_{MT}(\omega_m)&=& 
-2\cdot | \vec{A\, } |^2 \cdot \sum_{\epsilon_n,Q} \;
\langle \tilde \eta (P,Q) \underline G_2(P,Q) \rangle_{\vec p} \cdot
\langle \tilde \eta(P-W,Q) \underline G_2(P-W,Q) \rangle_{\vec p-\vec q'} \cdot L(Q), \\
\chi_{DOS}(\omega_m)&=& \quad 4\cdot | \vec{A\, } |^2 \cdot
\sum_{\epsilon_n,Q} \;
\langle \tilde \eta (P,Q)^2 \underline G_3(P,Q) \rangle_{\vec p} \, 
\langle  \underline G_1(P-W)  \rangle_{\vec p-\vec q'} \cdot L(Q)
 \, ,
\end{eqnarray}
and for the Pauli susceptibility,
\begin{eqnarray}
\chi_{MT+DOS}(0)&=& 8\cdot
\sum_{\epsilon_n,Q} \; \langle \tilde \eta (P,Q)^2 \underline G_4(P,Q)\rangle_{\vec p} \cdot L(Q),
\\
\chi_{AL}(0)&=& 8\cdot
\sum_{\epsilon_n,Q}  \Big( \langle \tilde \eta(P,Q) \, \underline G_3(P,Q)\rangle_{\vec p}\Big)^2
\cdot C(\epsilon_n,Q)\cdot  L(Q).
\end{eqnarray}
These results can be written more compactly
using the identities,
\begin{eqnarray}
&&\bigg\langle \tilde \eta (P,Q)^2 
\frac{\delta}{\delta \Sigma(\epsilon_n)} \underline G_2(P,Q) \bigg\rangle_{\vec p} = 
\frac{\delta}{\delta \Sigma(\epsilon_n)} 
\bigg\langle \eta (\vec p) \tilde \eta (P,Q) \underline G_2 (P,Q) \bigg\rangle_{\vec p} ,\\
&&\bigg\langle \tilde \eta (P,Q)^2 \frac{\delta^2}{\delta \Sigma(\epsilon_n)^2} \underline G_2(P,Q) \bigg\rangle_{\vec p} + 2 \; \bigg\langle \tilde \eta (P,Q)
\frac{\delta}{\delta \Sigma(\epsilon_n)} \underline G_2(P,Q) \bigg\rangle_{\vec p}^2 \; C(\epsilon_n,Q)
= 
\frac{\delta^2}{\delta \Sigma(\epsilon_n)^2} 
\bigg\langle \eta (\vec p) \tilde \eta (P,Q) \underline G_2 (P,Q) \bigg\rangle_{\vec p} .\\
\end{eqnarray}
Defining $G_1(\epsilon_n)= \Big\langle  \underline G_1(P) \Big\rangle_{\vec p}$,
$B_1(\epsilon_n,Q)=\Big\langle \tilde \eta (P,Q) \underline G_2 (P,Q) \Big\rangle_{\vec p}$
and $ 
B_2(\epsilon_n,Q)=\Big\langle \eta (\vec p) \tilde \eta (P,Q) \underline G_2 (P,Q) \Big\rangle_{\vec p}
$
we obtain for the NSLR rate,
\begin{eqnarray}
\label{wnmrMT}
\chi_{MT}(\omega_m)&=& 
-2\cdot | \vec{A\, } |^2 \cdot \sum_{\epsilon_n,Q} \; B_1(\epsilon_n,Q) \, B_1(\epsilon_n-\omega_m,Q)
 \cdot L(Q), \\
\label{wnmrDOS}
\chi_{DOS}(\omega_m)&=& \quad 4\cdot | \vec{A\, } |^2 \cdot
\sum_{\epsilon_n,Q} \; G_1(\epsilon_n-\omega_m) \frac{\delta}{\delta \Sigma(\epsilon_n)} B_2(\epsilon_n,Q)
\cdot L(Q)
 \, ,
\end{eqnarray}
and for the Pauli susceptibility,
\begin{eqnarray}
\label{wsus}
\chi_{MT+DOS+AL}(0)&=& 4\cdot
\sum_{\epsilon_n,Q} \; 
\frac{\delta^2}{\delta \Sigma(\epsilon_n)^2} B_2(\epsilon_n,Q)
\cdot L(Q).
\end{eqnarray}

\end{appendix}

\rrule
\bibliographystyle{unsrt}

\begin{thebibliography}{10}
\bibitem{anomMT}
The anomalous MT term (see Eq. (\ref{S_M}), last
line), is determined by $\left| B_1\right|^2$, which is equal to 
$\left| A_1\right|^2$ for d-wave pairing and $\left| A_0\right|^2$
for s-wave pairing in the clean limit. Because $\left| A_0\right|^2
\sim |(v_Fq)^2-(2\tilde\epsilon)^2|^{-1}$ 
the dominant contribution comes 
from the region $2\tilde \epsilon \approx v_F q$ 
in the clean limit for s-wave pairing. 
For this region from Eq. (\ref{A1}) in App. \ref{Afunctions} follows 
$\left| A_1(\epsilon,q,\phi) \right|^2 \approx 
2 (\cos 2\phi)^2 \left| A_0(\epsilon,q,\phi) \right|^2$, so that
in the clean limit the anomalous MT term for d-wave pairing
gives results similar
to the s-wave case,  i.e. 
a large positive contribution. The regular MT
term on the other hand contains the quantity $\left( A_1\right)^2$.
A principal-value integral leads to a result
which is proportional to $(v_Fq)^4$ for d-wave pairing and is negligible as can
be seen in Fig. \ref{F1d}.
With impurity scattering, the imaginary part of $\tilde \epsilon $
leads to suppression of the anomalous MT term.
\bibitem{LLQ}
In d-wave symmetry different Landau levels are coupled through higher
order terms in the pair momentum operator, $\vec q$.
This effect is
unimportant except possibly 
in very high fields near $H_{c2}(0)$.
In typical high-field experiments $H \lsim 30$ T
is still much lower than $H_{c2}(0)$.
To determine the mean field transition temperature in a finite magnetic
field we averaged $B_2 $ over $\phi $ for d-wave pairing to eliminate the
$\phi$-dependence in the fourth and higher-order terms in the momentum.
\bibitem{ALdiagram}
We use compact diagrams for our diagrammatic representation.
The standard way of drawing diagrams is obtained by resolving the
internal structure of the $K$-vertices and assigning explicit the
spin labels.
The set of four diagrams shown in Fig. \ref{low} is 
equivalent to the set of 10 diagrams given e.g. in ref. \onlinecite{varlamov98}.
We demonstrate this connection for the
compact form of the Aslamazov-Larkin diagram.
It contains for singlet pairing only one
pair propagator. The other collective mode is a Cooperon-type
impurity ladder in the particle-particle channel. One obtains 4 different
diagrams of this type.
In ref. \onlinecite{varlamov98} they are referred to also as
special types of Maki-Thompson and DOS diagrams
(counting only pairing fluctuation modes for
diagram naming schemes). The Aslamazov-Larkin diagram in the non-compact form
is exactly zero for spin singlet pairing.
The same is true for the corresponding contribution to the 
NSLR rate, but in this case it is additionally one order smaller in the 
parameter $T_c/E_F$.
\end{thebibliography}

\vspace{7cm}
\end{multicols}
\end{document}